\documentclass[acmsmall]{acmart}

\usepackage{newfloat}
\usepackage{listings}

\lstdefinelanguage{diff}{%
  morecomment=[f][\color{red}]{-},%
  morecomment=[f][\color{green!50!black}]{+},%
  morecomment=[f][\color{gray}]{@},%
}

\DeclareCaptionStyle{ruled}{labelfont=normalfont,labelsep=colon,strut=off} 
\floatstyle{ruled}
\newfloat{listing}{tb}{lst}{}
\floatname{listing}{Listing}

\usepackage{amsmath,amssymb,amsfonts}
\usepackage{graphicx}
\usepackage{textcomp}
\usepackage{amssymb}
\usepackage{minted}
 \usepackage{pgfplots} 
\pgfplotsset{compat=1.18}
\usepackage{multirow}
\usepackage{amsmath, balance}
\usepackage{scalerel}
\usepackage{tikz}
\usepackage{url}
\usepackage{academicons}
\usepackage{tabularray}  
\usepackage{lipsum}  
\usepackage{adjustbox}
\usepackage{mdframed}
\usepackage{orcidlink}
\usepackage{svg}
\usepackage{caption}
\usepackage{subcaption}
\usepackage{hyperref}

\usepackage{etoolbox}
\AtBeginEnvironment{thebibliography}{\catcode`\_=12\relax}

\usepackage{tcolorbox}
\usepackage[table,dvipsnames]{xcolor}

\usepackage{tabularx}
\usepackage{booktabs}
\usepackage{soul}

\usepackage{array}
\usepackage{longtable}
\usepackage{arydshln}
\usepackage{algorithm}
\usepackage{algpseudocode}

\AtBeginDocument{%
  }

\setcopyright{acmlicensed}  
\copyrightyear{2026}
\acmYear{2026}
\acmDOI{10.1145/3815588}
\acmJournal{TOSEM}

\newcommand{\JW}[1]{#1}  
\AtBeginDocument{\colorlet{blue}{black}}  

\begin{document}
\raggedbottom
\setlength{\emergencystretch}{3em}

\title{GrowthHacker: Automated Off-Policy Evaluation Optimization Using Code-Modifying LLM Agents
}


\author{Jie JW Wu\orcidlink{0000-0002-7895-2023}}
\authornote{Both authors contributed equally to this research.}
\affiliation{%
  \institution{Michigan Technological University, Houghton}
  \department{Department of Computer Science}
  \city{Houghton}
  \state{MI}
  \country{USA}
}
\email{jie.jw.wu@mtu.edu}

\author{Ayanda Patrick Herlihy\orcidlink{0009-0007-9565-0020}}
\authornotemark[1]
\affiliation{%
  \institution{Birmingham City University}
  \city{Birmingham}
  \country{UK}}
\email{ayanda.herlihy@mail.bcu.ac.uk}

\author{Ahmad Saleem Mirza\orcidlink{0009-0008-8685-2904}}
\authornote{Significant contribution.}
\affiliation{%
  \institution{University of British Columbia, Kelowna}
  \city{Kelowna}
  \country{Canada}}
\email{ahmadsm1@student.ubc.ca}

\author{Ali Afoud\orcidlink{0009-0006-7204-993X}}
\authornotemark[2]
\affiliation{%
  \institution{University of British Columbia, Kelowna}
  \city{Kelowna}
  \country{Canada}}
\email{afoudali@hotmail.com}

\author{Fatemeh Fard\orcidlink{0000-0002-4505-6257}}
\affiliation{%
  \institution{University of British Columbia, Kelowna}
  \city{Kelowna}
  \country{Canada}}
\email{fatemeh.fard@ubc.ca}



\begin{abstract}
\JW{With data-driven development now widely adopted in the software industry, online A/B testing is an established method for measuring the effects of new technologies. }However, deploying online experiments demands considerable resources for experimental design, implementation, and production deployment, and may negatively impact users (e.g., unsafe or unethical outcomes) while requiring multiple weeks of data collection.
To tackle these issues, a growing research area known as \textit{off-policy evaluation (OPE)} (offline A/B testing) has gained significant attention. Its goal is to assess new technologies offline by leveraging previously collected logged data. OPE is also a fundamental problem in reinforcement learning (RL) and is important in areas where online A/B testing is expensive or risky, such as healthcare, recommender systems, education, dialogue systems, and robotics.
Despite recent advances in code-generation large language models (LLMs) and agentic workflows, little is known about whether and how LLMs and LLM-based agents can be leveraged to automatically optimize OPE implementations.
In this work, we propose \textit{GrowthHacker}, a benchmark that evaluates baseline LLMs and LLM-based agents on large-scale public datasets. GrowthHacker autonomously and iteratively modifies code, runs OPE, and uses the resulting metrics to guide subsequent rounds of optimization. We evaluate methods on Open Bandit Pipeline (OBP)~\cite{saito2021openbanditdatasetpipeline} and Scope-RL~\cite{kiyohara2023scope}, and we develop a \textit{two\_agent} framework that addresses limitations of existing agent frameworks while reducing system complexity.
Across both libraries, \textit{two\_agent} demonstrates the highest reliability (98.1\%--100\% success rate) and the highest positive-outcome rate (78\%), with a median improvement of 4.4\% among positive outcomes; CrewAI achieves the highest average improvement (37.9\%) and is the only framework with zero extreme-value failures. AutoGen and Default each reach 65\% positive-outcome rates. These results establish the feasibility of using LLM-based agents as automated ``growth hackers’’ to continuously improve OPE systems, with implications for scaling data-driven decision-making in production environments where manual optimization is expensive.
Our replication package is available at \url{https://doi.org/10.5281/zenodo.17496869} and \url{https://github.com/jie-jw-wu/ope-agent}.
\end{abstract}

\keywords{Large Language Models, LLM-based Agents, Off-Policy Evaluation, Code Optimization, Automated Software Engineering, Reinforcement Learning, A/B Testing, Agentic AI}

\ccsdesc[500]{Computing methodologies~Reinforcement learning}
\ccsdesc[300]{Computing methodologies~Natural language processing}
\ccsdesc[500]{Software and its engineering~Software creation and management}
\ccsdesc[500]{Information systems~Retrieval models and ranking}

\settopmatter{printacmref=true}

\maketitle
\section{Introduction} \label{sec:introduction}
\JW{Many software organizations have adopted a data-driven culture, shifting from traditional requirement-based development to data-driven development and data-driven decision-making~\cite{auer2021controlled,fabijan2018online,wu2023comparison}. In this transition, \textit{online A/B testing}, often referred to as controlled online experiments, split testing, or randomized trials, is a standard method for assessing user behavior and the impact of product changes in web-based systems~\cite{fitzgerald2017continuous,fagerholm2017right,kohavi2020trustworthy,xu2015infrastructure}.} It is commonly applied in domains such as social media platforms~\cite{xu2015infrastructure}, search engines~\cite{tang2010overlapping}, online social networks~\cite{xu2015infrastructure,feitelson2013development}, and various web services~\cite{turnbull2019learning,pajkovic2022algorithms}. In this procedure, multiple versions of a feature or product are randomly assigned to different user segments, after which user interaction data is recorded and analyzed to compare performance against the original system~\cite{fitzgerald2017continuous,fagerholm2017right}. By experimenting on a relatively small subset of users, organizations can gather reliable evidence that helps stakeholders decide whether a new variant should be rolled out to the entire user base~\cite{kohavi2020trustworthy,xu2015infrastructure}.

However, online A/B testing suffers from several limitations. First, it takes significant development efforts to design and implement the change in the code base, with production-level standards. Second, the change will have a real impact on a relatively large group of users in the A/B test to get statistically significant A/B results. So it could affect users in a negative, unsafe, or unethical way if the change in the A/B test includes any bug or safety issue. Thus, domain owners of the products need to sign off for them to be served to a subset of users. Lastly, it typically needs several weeks to run the A/B tests to collect the data with potentially multiple iterations~\cite{kohavi2020trustworthy}. These limitations dramatically increase the time for the product team to test new ideas. 

To address these pain points, researchers have studied the emerging topic of \textit{Off-Policy Evaluation (OPE)} (or \textit{offline A/B testing}, \textit{counterfactual evaluation})~\cite{joachims2016counterfactual,gilotte2018offline,gruson2019offline,reklaite2022offline,saraswat2021hybrid}. The objective of offline A/B testing is to conduct offline evaluation of a new technology by estimating from historical logged data~\cite{joachims2016counterfactual}. A number of estimators have been developed such as Direct Method (DM)~\cite{beygelzimer2009offset}, Inverse Probability Weighting (IPW)~\cite{precup2000eligibility}, and Doubly Robust (DR)~\cite{dudik2014doubly} to reach a good balance between bias and variance~\cite{chen2023opportunities}, therefore increasing the correlation between the estimated results in offline A/B tests and the actual results in online A/B tests. 

More broadly, OPE is also a fundamental problem in reinforcement learning (RL)~\cite{uehara2022review}. Reinforcement learning algorithms optimize decision-making across domains like healthcare and recommender systems, generating valuable log data for evaluation. However, this log data is challenging to exploit because it only shows outcomes for selected actions and is biased toward the algorithm's preferences, unlike conventional supervised learning where all outcomes are observable. In applications where online A/B testing is expensive or risky, such as healthcare, recommender systems, education, dialog systems, and robotics, OPE becomes critical, especially for offline RL~\cite{uehara2022review}.

Although OPE is a promising approach due to much smaller development effort and faster turnaround time, there are still limitations for it to be reliably and effectively used in requirements engineering in practice. OPE is a manual process that runs the offline evaluation for manually selected algorithms (or policy~\cite{gilotte2018offline}) against the one-off historical data. Therefore, there is a lack of systematic updates of offline evaluation on code solutions or solution variants that could be potentially more optimal than the manually selected ones~\cite{wu2024autooffab}. This can lead to unreliable and less optimized results for OPE. To address this issue, a recent tool, \textit{pyIEOE (Interpretable Evaluation for Offline Evaluation)}, was proposed to conduct hyperparameter tuning of OPE estimators in the OBP dataset \cite{saito2021evaluating}. However, the tool is focused on tuning hyperparameters, and it needs users to input specifications about which hyperparameters to tune. In this research, we are interested in addressing this issue by optimizing the code space rather than selecting and tuning hyperparameters manually.

\begin{figure}[t]
\begin{lstlisting}[language=diff, basicstyle=\scriptsize\ttfamily, numbers=left, numbersep=3pt, xleftmargin=1.5em, escapeinside={(*}{*)}]
 ipw_lr = IPWLearner(
     n_actions=dataset.n_actions,
(*\colorbox{red!20}{-\ \ \ \ base\_classifier=LogisticRegression(C=1, random\_state=12345)}*)
(*\colorbox{green!20}{+\ \ \ \ \# Suggested Change 1: Increase C for LogisticRegression in IPWLearner}*)
(*\colorbox{green!20}{+\ \ \ \ base\_classifier=LogisticRegression(C=10, random\_state=12345)}*)
 )
 ipw_lr.fit(
     context=bandit_feedback_train["context"],
     ...
 )
 action_dist_ipw_lr = ipw_lr.predict(context=bandit_feedback_test["context"])
 ipw_rf = IPWLearner(
     n_actions=dataset.n_actions,
(*\colorbox{red!20}{-\ \ \ \ base\_classifier=RandomForest(n\_estimators=15, min\_samples\_leaf=100, random\_state=12345)}*)
(*\colorbox{green!20}{+\ \ \ \ \# Suggested Change 2: Decrease min\_samples\_leaf for RandomForest in IPWLearner}*)
(*\colorbox{green!20}{+\ \ \ \ base\_classifier=RandomForest(n\_estimators=15, min\_samples\_leaf=20, random\_state=12345)}*)
 )
 ipw_rf.fit(
     context=bandit_feedback_train["context"],
     ...
 )
 action_dist_ipw_rf = ipw_rf.predict(context=bandit_feedback_test["context"])
 
 regression_model = RegressionModel(
     n_actions=dataset.n_actions,
     action_context=dataset.action_context,
(*\colorbox{red!20}{-\ \ \ \ base\_model=LogisticRegression(random\_state=12345),}*)
(*\colorbox{green!20}{+\ \ \ \ \# Suggested Change 3a: Add C for LogisticRegression within RegressionModel}*)
(*\colorbox{green!20}{+\ \ \ \ base\_model=LogisticRegression(C=10, random\_state=12345),}*)
 )
 estimated_rewards_by_reg_model = regression_model.fit_predict(
     context=bandit_feedback_test["context"],
     action=bandit_feedback_test["action"],
     reward=bandit_feedback_test["reward"],
(*\colorbox{red!20}{-\ \ \ \ n\_folds=3, \# use 3-fold cross-fitting}*)
(*\colorbox{green!20}{+\ \ \ \ \# Suggested Change 3b: Increase n\_folds for cross-fitting}*)
(*\colorbox{green!20}{+\ \ \ \ n\_folds=5,}*)
     random_state=12345,
 )
 ope = OffPolicyEvaluation(
\end{lstlisting}
\caption{\JW{Successful motivating example of automated OPE code optimization for data-driven requirements validation. Red lines (--) show baseline hyperparameters; green lines (++) show optimizations by Two-Agent system. Changes include: regularization strength (\texttt{C}: 1$\to$10), ensemble leaf size (\texttt{min\_samples\_leaf}: 100$\to$20), added regularization to regression model, and cross-validation folds (\texttt{n\_folds}: 3$\to$5). After applying these changes, the optimized relative\_ee\_ipw\_lr metric value is 0.1635 for estimator DM.}}
\Description{A side-by-side code-diff style illustration showing baseline OPE hyperparameters in red and optimized hyperparameters in green, highlighting changes to regularization, ensemble leaf size, regression regularization, and cross-validation folds.}
\label{fig:ope-diff}
\end{figure}

\JW{This work addresses a software engineering challenge at the intersection of requirements engineering and data-driven development: validating and refining data-driven requirements when online experimentation is prohibitively expensive. In modern software systems, requirements for recommendation engines, search algorithms, and personalization features are increasingly data-driven rather than specification-based~\cite{wu2024autooffab}, yet validation via online A/B testing imposes substantial overhead (e.g., production-quality implementation, prolonged experimentation, and exposure to real users)~\cite{fabijan2018online}. Off-policy evaluation (OPE) offers an alternative, but current practice remains manual and expertise-intensive, limiting its use in continuous development workflows.}

\JW{From a software engineering perspective, the core challenge is scalability: how can teams systematically explore and validate multiple solution variants without repeatedly deploying them online? This setting is analogous to continuous integration and automated testing, except that ``correctness'' is assessed through statistical estimation rather than deterministic test outcomes. We study whether LLM-based agents can automate parts of this validation loop by generating and evaluating constrained code variants. Figure~\ref{fig:ope-diff} illustrates a representative example, where the agents propose simple, localized modifications (e.g., \texttt{C}: 1$\to$10, \texttt{min\_samples\_leaf}: 100$\to$20, \texttt{n\_folds}: 3$\to$5) without manual tuning.}

On the other hand, given the recent advances in Large Language Models (LLM) and agentic flow for code generation, the LLM and agent approaches for coding optimization are receiving increasing attention. For example, RAPGen \cite{mao2023rapgen} utilizes OpenAI Codex \cite{chen2021evaluating} in a zero-shot setting to address inefficiencies in C\# code, achieving high precision without requiring additional training. SBLLM~\cite{gao2024search} is a search-based LLM framework to enable code optimizations. However, little has been explored on connecting LLM or agent-based code optimization to automating data-driven decisions such as OPE, which could potentially lead to high benefits in both academia and industry. In academia, OPE is heavily used in reinforcement learning, and thus automating the optimization of OPE via code refinement reveals an early signal of autonomous machine programming toward machine self-evolution. In industry, OPE is directly tied to key business objectives and performance metrics, including growth, revenue, as well as other considerations such as safety and privacy. Automating the optimization of these metrics is therefore highly valuable. In this work, our goal is to answer the question of whether LLM or agent-based code optimization can serve as a silent ``growth hacker'' behind the scenes to automatically and effectively improve OPE results. If this approach can be explored to create more intelligent non-trivial code changes for OPE and achieve more performant results, it would demonstrate significant potential for automating OPE by optimizing directly in code space, rather than relying solely on hyperparameter search-based methods~\cite{wu2023multi}. \JW{However, as shown in the experiment section, current agent approaches on OPE face limitations, including demonstrated compilation and runtime failures, tool utilization inefficiencies, and context degradation across iterations, as further discussed in Section~\ref{sec:limitation}.}

\begin{figure*}
    \centering
    \includegraphics[width=1\textwidth]{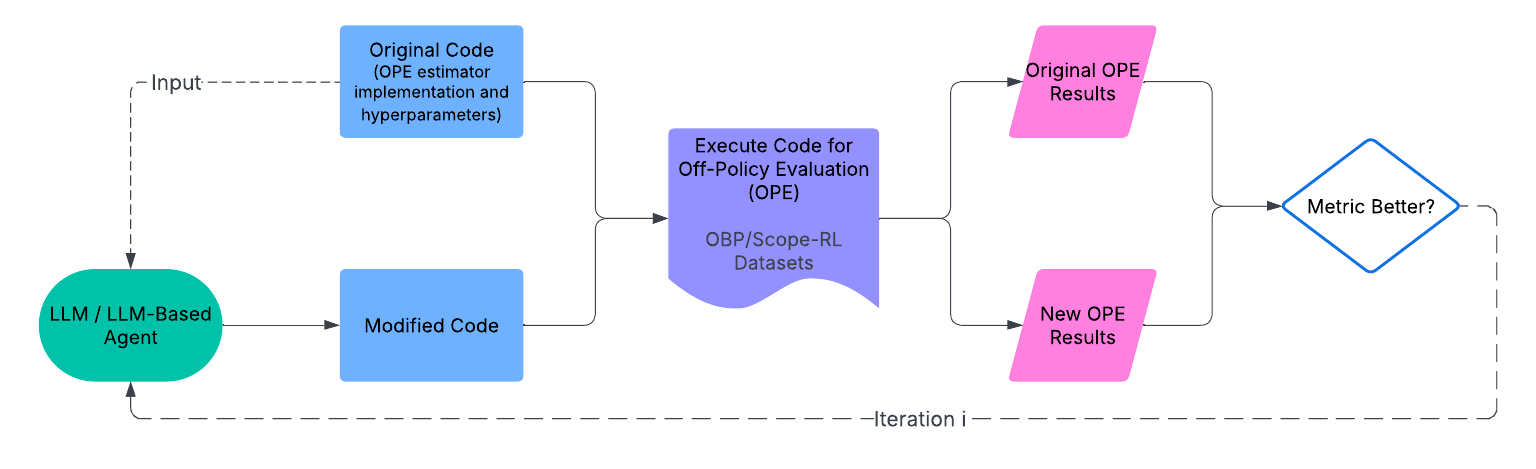}
    \caption{Flow Diagram for GrowthHacker, a benchmark system for optimizing Off-Policy Evaluation through code modification, using a LLM or LLM agent.}
    \Description{A workflow diagram showing GrowthHacker converting a notebook to Python, generating candidate code changes with an LLM or agent, executing the modified code, collecting OPE metrics, and selecting the best iteration.}
    \label{fig:growth_hacker}
\end{figure*}

In this work, we propose \JW{\textit{GrowthHacker}}, the first benchmark system for optimizing OPE through code modification, using an LLM or LLM-based agents. In the GrowthHacker system, the LLM or LLM-based agent autonomously and iteratively optimizes code, applies the code, gets OPE results, and compares them to the original OPE results. Figure~\ref{fig:growth_hacker} shows the flow diagram of the GrowthHacker benchmark system. First, we collected the dataset, designed and set up the experiment protocols. Second, we created baselines of LLMs and LLM-based agents to evaluate and analyze the performance of these methods for optimizing OPE on Open Bandit Pipeline (OBP) dataset~\cite{saito2021openbanditdatasetpipeline} and Scope-RL datasets~\cite{kiyohara2023scope}. Third, we designed the \textit{two\_agent} framework that addresses limitations in the existing agent framework, reduces system complexity while maintaining optimization effectiveness. 

Our experimental results demonstrate that LLM-based agents can achieve significant improvements in OPE performance through code modification. The \texttt{two\_agent} framework demonstrates the highest reliability (98.1\%--100\% success rate) and the highest positive-outcome rate (78\%), with a median improvement of 4.4\% among positive outcomes; CrewAI achieves the highest average improvement (37.9\%) and is the only framework with zero extreme-value failures. AutoGen and Default each reach 65\% positive-outcome rates. These results establish the feasibility of using LLM-based agents as automated ``growth hackers'' for continuously improving OPE systems, with implications for scaling data-driven development activities such as growth hacking and data-driven decision-making in production environments where manual optimization is expensive.

Our contributions are summarized as follows:
\begin{itemize}
\item We propose GrowthHacker, the first benchmark system for evaluating the performance of optimizing OPE via code modifications. We conducted a comprehensive benchmarking to evaluate the performance in automated code optimization of OPE for LLMs and LLM-based agents. 
\item We proposed a new two\_agent framework using one Analyzer and the other a Code Generator for improving the OPE results. The two\_agent framework records the lowest failure rate (0\%--1.9\%) and highest success rate (98.1\%--100\%), with superior optimization performance and the highest median improvement (4.4\%) among positive outcomes.
\item We release full code for full replication and future research on GrowthHacker, on Zenodo (\url{https://doi.org/10.5281/zenodo.17496869}) and GitHub (\url{https://github.com/jie-jw-wu/ope-agent}).
\end{itemize}

\section{Background}
 
\subsection{The Off-Policy Evaluation Background}
Following the notations in~\cite{saito2021evaluating}, in the contextual bandit problem, a learner observes contexts $x \in \mathcal{X}$ (e.g., user characteristics), selects actions $a \in \mathcal{A}$ from a finite set, and receives bounded rewards $r \in [0, r_{\max}]$ (e.g., conversion outcomes). The environment is characterized by unknown distributions $p(x)$ for contexts and $p(r|x,a)$ for rewards given context-action pairs. A policy $\pi: \mathcal{X} \rightarrow \Delta(\mathcal{A})$ assigns probabilities $\pi(a|x)$ to actions conditioned on contexts. The expected reward function is defined as $q(x,a) = \mathbb{E}[r|x,a]$. We observe a dataset $\mathcal{D} = \{(x_i, a_i, r_i)\}_{i=1}^n$ generated by the behavior policy $\pi_b$, where each triplet is sampled independently according to
\[
\prod_{i=1}^n p(x_i)\, \pi_b(a_i \mid x_i)\, p(r_i \mid x_i, a_i).
\]

The objective is to evaluate a target policy $\pi_e \neq \pi_b$ by estimating its expected value:

\begin{equation}
V(\pi_e) = \mathbb{E}_{p(x)\pi_e(a|x)p(r|x,a)}[r],
\end{equation}

without deploying it online. This off-policy estimation problem is fundamental in applications where online experimentation is costly or risky.

\textbf{Existing OPE Estimators.} Off-policy evaluation relies on several estimators to approximate the policy value $V(\pi_e)$ from logged data $\mathcal{D}$. We describe three of them. Please refer to \cite{saito2021openbanditdatasetpipeline} for a more complete list of estimators. The Direct Method (DM)~\cite{beygelzimer2009offset} fits an estimate of the reward function, denoted by $\hat{q}(x,a)$, and evaluates the target policy by averaging the predicted rewards under $\pi_e$:
\[
\widehat{V}_{\mathrm{DM}}(\pi_e; \mathcal{D}, \hat{q})
= \mathbb{E}_n \Big[\, \mathbb{E}_{a \sim \pi_e(\cdot \mid x_i)} \big[\hat{q}(x_i, a)\big] \Big].
\]
While simple to implement, the quality of DM depends heavily on how well $\hat{q}$ approximates the true reward function.

Inverse Probability Weighting (IPW)~\cite{precup2000eligibility} instead reweights the observed rewards using the importance ratio
\[
\rho(x,a) = \frac{\pi_e(a \mid x)}{\pi_b(a \mid x)},
\]
leading to the estimator
\[
\widehat{V}_{\mathrm{IPW}}(\pi_e; \mathcal{D})
= \mathbb{E}_n \big[ \rho(x_i, a_i)\, r_i \big].
\]
Although unbiased when $\pi_b$ is exactly known, IPW may suffer from large variance if $\pi_e$ differs significantly from $\pi_b$.

To mitigate this issue, the Doubly Robust (DR) estimator~\cite{dudik2014doubly} augments IPW with the reward model:
\[
\widehat{V}_{\mathrm{DR}}(\pi_e; \mathcal{D}, \hat{q}) =
\mathbb{E}_n \left[
\mathbb{E}_{a \sim \pi_e(\cdot \mid x_i)}[\hat{q}(x_i, a)]
+ \rho(x_i,a_i)\big(r_i - \hat{q}(x_i,a_i)\big)
\right],
\]
which uses $\hat{q}$ as a variance-reducing control variate and remains unbiased if either the model or the behavior policy is correctly specified.

\textbf{Evaluating Offline Evaluation.} The standard metric for assessing OPE estimator performance is mean squared error, defined as
\[
\text{MSE}(\hat{V}; \pi_e, \theta) := \mathbb{E}_{\mathcal{D}}\big[(V(\pi_e) - \hat{V}(\pi_e; \mathcal{D}, \theta))^2\big].
\]
This quantity corresponds to the expected squared deviation between the true policy value and the estimator's output, where a smaller MSE indicates better performance.



\subsection{Problem Formulation}

We formulate the LLM-based code optimization task as follows: Given source code $\mathcal{C}$ implementing OPE with initial parameters $\theta_0$, historical dataset $\mathcal{D}$, and OPE performance metrics (e.g., relative estimation error for each estimator), the agent must generate a sequence of code modifications $\{(\mathcal{C}_t, \theta_t)\}_{t=1}^T$ that minimizes the relative estimation error, such that $\min_{t \in \{1,\ldots,T\}}\,\epsilon(\mathcal{C}_t,\theta_t) < \epsilon(\mathcal{C}_0,\theta_0)$, while preserving functional correctness of the implementation. Note that in this work, the parameters $\theta_t$ are included as part of the source code $\mathcal{C}_t$.

\section{The Two-Agent Framework}
\begin{figure*}
    \centering
    \includegraphics[width=1\textwidth]{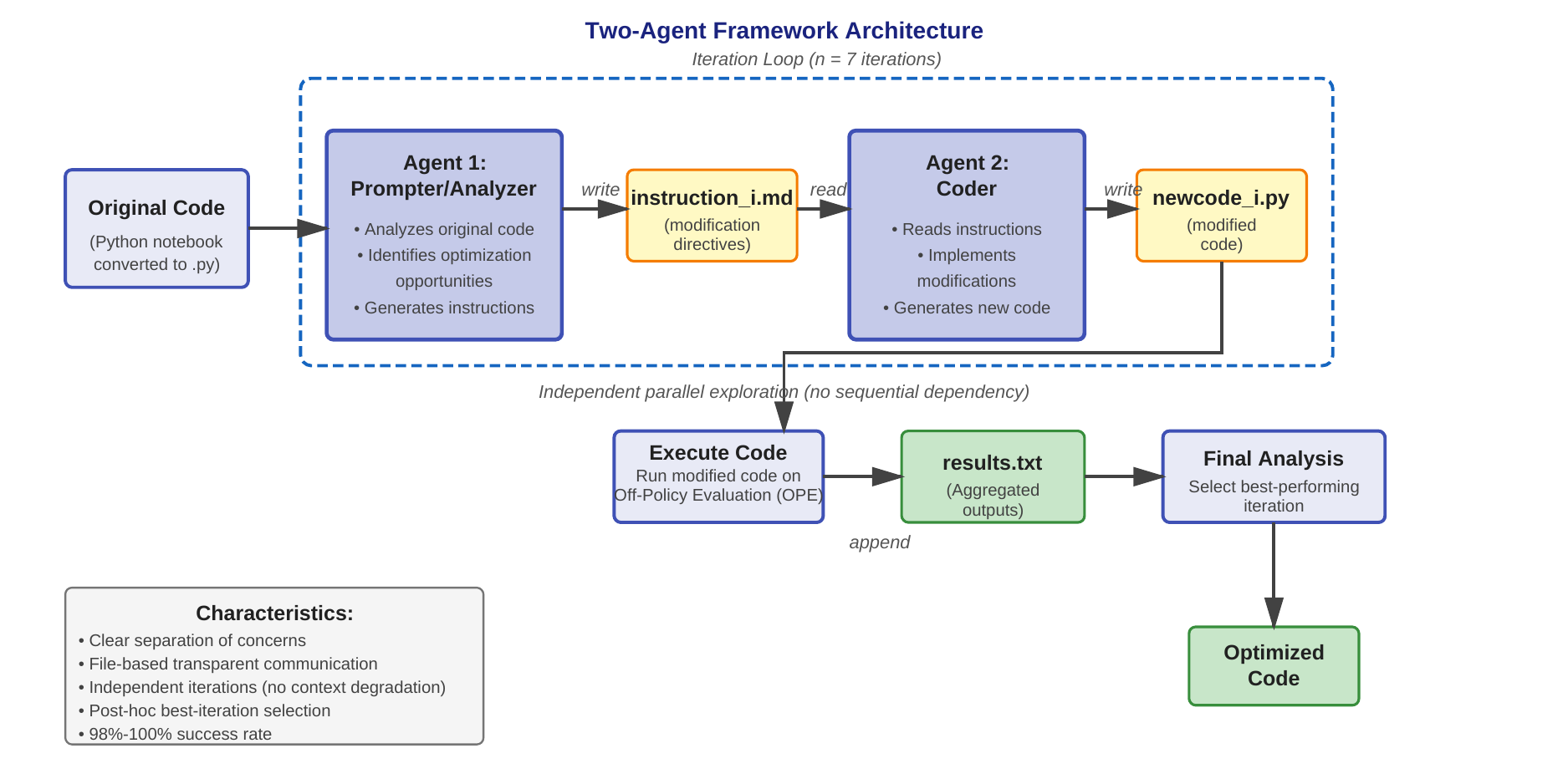}
    \caption{Visual Illustration for Two-Agent Architecture}
    \Description{A two-agent architecture diagram showing an Analyzer Agent writing structured instruction files, a Coder Agent reading those instructions to generate code variants, and an execution stage that evaluates each variant and records results.}
    \label{fig:final_design}
\end{figure*}

As shown in Figure~\ref{fig:final_design}, the Two-Agent framework operates through an iterative process: (1) the LLM agent analyzes the current code structure and parameters, (2) proposes modifications based on domain knowledge, (3) executes the modified code to compute performance metrics in OPE, 
 and (4) repeats for $T$ iterations. This approach transforms the traditionally manual hyperparameter tuning process into an automated code optimization process.

\subsection{Limitations of Existing Agent Architectures}
\label{sec:limitation}
CrewAI and AutoGen frameworks exhibit fundamental limitations when applied to OPE optimization tasks, as confirmed by the evaluation in Section~\ref{sec:results} (Tables~\ref{tab:failure_analysis} and~\ref{tab:failure_types}). Three critical challenges emerged:

\textbf{Compilation Failures}: Both CrewAI and AutoGen exhibited persistent compilation issues due to deprecated syntax generation and incorrect API usage. \textcolor{blue}{The formal evaluation reports failure rates of 9.9\% (CrewAI) and 7.4\% (AutoGen) overall (Table~\ref{tab:failure_analysis}), with syntax and code errors accounting for 62\% of all failures across frameworks (Table~\ref{tab:failure_types}).} These failures stemmed from the frameworks' inability to maintain syntactic consistency when generating complex code modifications\JW{~\cite{wang2025towards}}.

\textbf{Tool Utilization Inefficiency}: While CrewAI demonstrated strong contextual understanding in text-based analysis, it struggled with tool utilization in code generation tasks. Agents frequently misused file operations, ignored specific prompts, or hallucinated function signatures, particularly when handling OPE library-specific APIs~\JW{\cite{lin2025llm,qin2023toolllm}}. \textcolor{blue}{File corruption caused by diff-syntax injection into generated Python code was the second most common failure pattern (18\% of failures), predominantly in AutoGen (Table~\ref{tab:failure_types}).}

\textbf{Context Degradation}: Both frameworks exhibited context degradation over multiple iterations. Agents failed to build upon previous attempts effectively, often repeating unsuccessful strategies or losing track of successful parameter modifications across iterations\JW{~\cite{cemri2025multi}}. \textcolor{blue}{This observation directly motivated the independent-iteration design of two\_agent, in which each iteration starts from the original baseline code rather than the previous iteration's output.}

\subsection{Simplified Two-Agent Architecture}
In response to the limitations of complex multi-agent frameworks, we developed a streamlined two-agent architecture that reduces system complexity while maintaining optimization effectiveness. The framework comprises two specialized agents with clearly delineated responsibilities:

\textbf{Prompter/Analyzer Agent} examines the original code and its execution results, identifies strengths and weaknesses, and generates detailed modification instructions. The agent focuses on suggesting simple, straightforward changes---primarily hyperparameter tuning and minor function adjustments---that are readily executable by an LLM.

\textbf{Coder Agent} receives the instruction file along with the original code, implements the specified modifications, and produces syntactically correct, functional code.

This separation of concerns addresses the compilation failure issues observed in previous approaches by limiting each agent's cognitive load and task complexity, thereby preventing the errors that emerged in more sophisticated multi-agent systems.

\subsection{Implementation Details}
The two-agent framework implements a structured workflow with explicit file-based communication that ensures transparency and enables systematic debugging. The architecture maintains the following file structure:

\begin{itemize}
\item \texttt{instruction\_i.md} contains modification directives generated by the Analyzer Agent for iteration $i$
\item \texttt{newcode\_i.py} stores the modified code produced by the Coder Agent based on \texttt{instruction\_i.md}
\item \texttt{result.txt} contains a comprehensive results table generated after all iterations complete, enabling post-hoc analysis of optimization trajectories
\end{itemize}

Algorithm~\ref{alg:two-agent-workflow} shows the pseudocode for the workflow. The optimization workflow proceeds as follows:
\begin{enumerate}
\item \textbf{Initial Setup}: The source notebook is converted to a Python file and executed via the \texttt{run\_code} script, with output saved to \texttt{result.txt}.

\item \textbf{Iteration Loop}: For each of $n$ iterations (typically $n=7$):
\begin{enumerate}
\item The Analyzer Agent retrieves the original code
\item It identifies optimization opportunities and generates modification instructions, stored in \texttt{instruction\_i.md}
\item The Coder Agent reads \texttt{instruction\_i.md} and the original code, implements the specified changes, and saves the result as \texttt{newcode\_i.py}
\item The \texttt{run\_code} script executes the modified code and saves the resulting OPE metrics
\end{enumerate}

\item \textbf{Final Analysis}: Finally, we use a lightweight LLM combined with a weighted combination of OPE estimator metrics to analyze all iteration results and select the best-performing configuration. This automated selection is effective in selecting the best iteration from the heterogeneous output formats across different notebooks.
\end{enumerate}

This file-based architecture provides complete transparency and debugging capabilities absent in black-box agent frameworks, enabling systematic analysis of each optimization step.

\begin{algorithm}
\caption{\JW{Two-Agent Framework Workflow}}
\label{alg:two-agent-workflow}
\begin{algorithmic}[1]
\State \JW{\textbf{Input:} Source notebook, number of iterations $n$ (default $n=7$)}
\State \JW{\textbf{Output:} Best-performing code variant and its OPE metrics}
\State \JW{\textcolor{gray}{// \textbf{Initial Setup}}}
\State \JW{$originalCode \gets$ \Call{ConvertNotebookToPython}{sourceNotebook}}
\State \JW{$baselineResults \gets$ \Call{ExecuteCode}{$originalCode$}}
\State
\State \JW{\textcolor{gray}{// \textbf{Iterative Optimization via Two Agents}}}
\Function{TwoAgentIteration}{\JW{$i$, $originalCode$}}
    \State \JW{\textcolor{gray}{// Analyzer Agent: Generate optimization instructions}}
    \State \JW{$instructionFile \gets$ \texttt{instruction\_i.md}}
    \State \JW{\Call{AnalyzeAndWriteInstructions}{$originalCode$, $instructionFile$}}
    \State \JW{\textbf{assert} \Call{FileExists}{$instructionFile$} \textcolor{gray}{// Verify file-based communication}}

    \State \JW{\textcolor{gray}{// Coder Agent: Implement modifications}}
    \State \JW{$newCodeFile \gets$ \texttt{newcode\_i.py}}
    \State \JW{\Call{ReadAndGenerateCode}{$instructionFile$, $originalCode$, $newCodeFile$}}

    \State \JW{\textcolor{gray}{// Execute and save OPE metrics}}
    \State \JW{\Call{ExecuteCode}{$newCodeFile$}}
\EndFunction
\For{\JW{$i = 1$ to $n$}}
    \State \JW{\Call{TwoAgentIteration}{$i$, $originalCode$}}
\EndFor
\State
\State \JW{\textcolor{gray}{// \textbf{Final Analysis}: Select best iteration}}
\State \JW{$(bestIteration, bestResults) \gets$ \Call{SelectBestFromResults}{$baselineResults$}} 
\State \JW{$bestCode \gets$ \texttt{newcode\_$bestIteration$.py}}
\State \JW{\Return $(bestCode, bestResults)$}
\end{algorithmic}
\end{algorithm}

\JW{\textbf{Intermediate Files as Non-LLM Control Layers}: As shown in Algorithm~\ref{alg:two-agent-workflow}, the two\_agent framework adopts a file-based serialization approach in which agents communicate exclusively through structured intermediate files. These files collectively form a non-LLM control layer that operates independently of LLM behavior. The workflow is organized around three files: \texttt{instruction\_i.md}, which contains structured markdown directives and serves as the sole communication channel between agents; \texttt{newcode\_i.py}, which stores the code generated by the Coder Agent after reading both the instruction file and the baseline code; and \texttt{result.txt}, which aggregates outputs from all iterations for post-hoc analysis.}

\JW{In contrast to multi-agent systems such as AutoGen and CrewAI, where agents coordinate through conversational protocols, the proposed design relies on one-way file-based communication. The Analyzer Agent writes \texttt{instruction\_i.md} without expecting interactive feedback, and the Coder Agent subsequently reads this file as a unidirectional input. This design removes the need for back-and-forth negotiation and reduces coordination overhead associated with task negotiation, context sharing, and state synchronization via LLM-mediated dialogue. The strict schema enforced for each file (markdown instructions, executable Python code, and structured results) enables early validation, allowing schema violations to be detected before execution and limiting error propagation. Additionally, each of the seven iterations operates independently from the same baseline code, avoiding sequential dependencies and preventing context accumulation across iterations. This simplicity is an intentional design decision aimed at improving robustness by avoiding failure modes commonly observed in more complex multi-agent architectures.}

\subsection{Iterative Optimization Strategy}
Unlike single-shot optimization approaches that commit to modifications without validation, our two-agent framework implements a multi-iteration exploration strategy with post-hoc selection. This approach offers several advantages:

\textbf{Multi-Iteration Exploration}: The system conducts 7 independent optimization attempts within each run, each deriving insights from the original code rather than previous iterations. Across iterations, we additionally apply novelty checking to avoid reusing previously tried parameter settings. This multi-iteration exploration strategy avoids the context degradation and error accumulation observed in sequential multi-agent systems.

\textbf{Best-Iteration Selection}: Rather than defaulting to the final iteration, a lightweight LLM combined with a weighted combination of OPE estimator metrics evaluates all iterations against the baseline and selects the configuration achieving optimal OPE metrics. \textcolor{blue}{In our experiments, this explicit post-hoc selection step was  helpful for two\_agent to succeed.}

\textbf{Targeted Optimization Scope}: The framework's prompts specify that code optimizations should focus primarily on hyperparameter tuning and minor function adjustments---modifications that current LLMs can reliably execute. In our debugging of experiments, this prompt design enables consistent compilation success while avoiding the syntactic errors and deprecated function calls that affected more untargeted code generation approaches in other baseline work (CrewAI and AutoGen).
However, more broadly, we should note that choosing between creative code-level changes and pure hyperparameter tuning is non-trivial due to the exploration--exploitation tradeoff.

 \textcolor{blue}{\textbf{Default configuration choice.} During development, we found that two\_agent is most effective when (i) applying modifications at the \texttt{whole\_code} level and (ii) using a larger iteration budget. This may be due to aspects of two\_agent's design, such as post-hoc selection of the best-performing iteration and novelty checking to avoid previously tried parameter settings. As mentioned earlier, choosing between creative code-level changes and pure hyperparameter tuning is non-trivial due to the exploration--exploitation tradeoff. In our experiments, we observe higher failure rates when we allow LLMs to explore more aggressively (i.e., propose broader code edits); therefore, our default prompts bias the agent toward hyperparameter tuning and minor, low-risk edits. We emphasize that this default choice reflects our experimental setting and should not be interpreted as a universal recommendation. Unless otherwise specified, we report two\_agent under this configuration; for cross-framework comparison, we also run a fully balanced setting that evaluates two\_agent across all modification methods under the same protocol as other frameworks (reported in Section~\ref{sec:results}).}

\section{GrowthHacker Experimental Design}
In this section, we first describe the research questions. Then we describe the experimental setup and the experiment results, followed by discussions and recommendations of the results.


\subsection{Research Questions}

We formulate four primary research questions to guide our experimental investigation:

\textbf{RQ1: What are the success rates and failure patterns when agent frameworks and standalone LLMs are applied to optimizing code for offline policy evaluation?} This question examines the reliability and failure modes of different approaches, providing insights into the practical challenges of automated code optimization. Understanding both success rates and the nature of failures is crucial for developing robust optimization strategies.

\textbf{RQ2: What is the comparative performance of the two\_agent framework and other agentic frameworks (AutoGen, CrewAI) in code optimization tasks for OPE?} Given the architectural differences between these frameworks, we seek to understand their relative strengths and weaknesses in the context of OPE optimization.

\textbf{RQ3: Which code modification approach (whole code generation, manual patch, or agent-applied patch) yields the most reliable optimizations?} The method of applying code changes significantly impacts success rates and optimization quality. We compare three distinct approaches: generating complete optimized files, creating and applying patches using Unix's patch command, and having LLM-based agents apply patches programmatically.

\textcolor{blue}{\textbf{RQ4: How does the two\_agent framework compare with traditional (non-LLM) hyperparameter optimization, and how do different prompt variants (hyperparameter-focused vs.\ code-level) affect optimization outcomes?} This question examines whether LLM-based optimization provides added value over conventional HPO baselines and explores the distinct contributions of parameter tuning versus code-level modifications to overall optimization performance.}


\subsection{Experimental Setup}

\subsubsection{Overview and Scale}

The experimental infrastructure was carefully designed to ensure fair comparison across frameworks while capturing detailed performance metrics. Each experimental run was monitored for success or failure 
, with comprehensive logging of error patterns, performance improvements, and computational characteristics. A failed run indicates the tool did not fulfill its core function: generating a completed execution with results of OPE that can be analyzed in the output CSV file.
  The use of Gemini-2.5-Flash as the underlying language model was chosen based on its balance of performance and cost-effectiveness, though the framework is designed to accommodate any LLM through standardized APIs.

\textcolor{blue}{\subsubsection{Two\_agent configuration and balanced evaluation.}
The two\_agent framework was developed and empirically tested in-house. During development, we found that combining the \texttt{whole\_code} modification method with a higher iteration budget (7 independent iterations) consistently yielded the strongest results. This advantage is driven by two design features: (i) across iterations, two\_agent tracks previously used hyperparameter values to avoid repeatedly proposing the same settings; and (ii) two\_agent includes a selection step that chooses the best-performing iteration post hoc, unlike the other LLM frameworks. Accordingly, our initial experiments evaluated two\_agent only under this single configuration (7 iterations, whole\_code), while the other frameworks were exercised across all three modification methods; we acknowledge that this introduced an imbalance in the cross-framework comparison. To ensure a fair comparison, we re-ran two\_agent under the same experimental protocol as the other frameworks (all three modification methods, the same 18 notebooks, and the same number of repetitions). The full evaluation follows a balanced factorial design (18 notebooks $\times$ 3 repetitions $\times$ 4 frameworks $\times$ 3 methods = 648 runs; 216 per method; 162 per framework), devoting equal numbers of runs to \texttt{two\_agent (equal)} and to each competing framework. For this balanced comparison, two\_agent was configured to use one internal iteration per outer repetition (matching the other frameworks), with its write-back mechanism ensuring each outer repetition builds sequentially on the previous result. We note that when deployed as a standalone tool (rather than as a direct competitor in a balanced cross-framework evaluation), the original 7-iteration configuration with post-hoc best-iteration selection is expected to yield stronger outcomes; the qualitative analysis in Section~5.5 illustrates this design in action.}

\subsubsection{Testing Configuration}

The testing parameters were systematically varied to explore the full optimization space:

\begin{itemize}
    \item \textbf{Frameworks Evaluated}: We tested four distinct approaches: (1) Default - standalone LLM implementation without agentic capabilities, (2) AutoGen - Microsoft's multi-agent framework (for \texttt{agent\_applies}: Suggester + Applier agents via conversational protocol; for \texttt{whole\_code} and \texttt{manual\_patch}: single AssistantAgent call), (3) CrewAI - an open-source multi-agent platform (for \texttt{agent\_applies}: Reader + Applier agents in a sequential Crew; for \texttt{whole\_code} and \texttt{manual\_patch}: single-agent Crew), and (4) two\_agent - a custom dual-agent architecture with internal iteration cycles.
    
    \item \textbf{Code Modification Modes}: Three code modification strategies were evaluated: whole\_code generation (complete file regeneration), manual\_patch (generating patches for Unix patch command), and agent\_applies (programmatic patch application by agents).
    
    \item \textbf{Iteration Strategy}: All four frameworks use the same outer repetition counts (n=1, 2, or 3), with each round applying one sequential modification step that builds on the previous result (each framework writes its modified code back to the input file before the next round begins). The best outer-round result across all repetitions is reported.
\end{itemize}

\subsubsection{Experimental Datasets}

The selection of evaluation notebooks was designed to provide comprehensive coverage of offline policy evaluation scenarios while presenting varying levels of optimization complexity. We selected datasets from Open Bandit Pipeline (OBP) and Scope-RL based on: (1) widespread adoption in OPE research~\cite{saito2021openbanditdatasetpipeline, kiyohara2023scope}, (2) production-level data quality with real-world datasets like Zozotown e-commerce data, (3) varied complexity from synthetic experiments to production-scale scenarios, and (4) diverse action spaces (discrete and continuous). OBP notebooks (200-500 LOC) provide focused evaluation with three core estimators (DM, IPW, DR), while Scope-RL notebooks (400-1,500 LOC) present greater complexity with 15+ estimators across three domain families (Basic, REC, RTB) with varying difficulty levels (Advanced/Basic/Zoo). This diversity ensures our findings have a certain level of generalizability for automated code optimization scenarios in practice.

For the OBP dataset~\cite{saito2021openbanditdatasetpipeline}, we utilized three carefully selected Jupyter notebooks that represent different aspects of bandit learning: The \texttt{multiclass.ipynb} notebook implements multi-class classification within a contextual bandit framework, providing insights into how optimization strategies handle discrete action spaces with multiple classes. The \texttt{obd.ipynb} notebook works with the complete Zozotown Research dataset, containing real-world fashion e-commerce data including product positions, anonymized features, and user click behavior. This notebook presents the challenge of optimizing estimators on production-scale data. The \texttt{synthetic.ipynb} notebook uses synthetically generated datasets, allowing controlled experimentation with known ground truth values for more precise performance measurement.

For Scope-RL dataset~\cite{kiyohara2023scope}, our evaluation spans 15 notebooks organized into three distinct families, each representing different reinforcement learning contexts. The Basic family consists of 6 notebooks covering fundamental RL scenarios across both continuous and discrete action spaces, with variations in complexity (Advanced/Basic/Zoo). The REC (Recommendation) family includes 3 notebooks focused on discrete-action recommendation scenarios, while the RTB (Real-Time Bidding) family comprises 6 notebooks addressing bidding optimization in both continuous and discrete settings. This comprehensive coverage ensures our findings generalize across diverse OPE applications.

\textbf{Bypass logic for Scope-RL.} We implemented a bypass logic within the test runners for Scope-RL. This logic detects computationally expensive training calls (e.g., model.fit() or dataset.obtain\_episodes()) in the notebook code. The reason is that, for all of the 15 Scope-RL notebooks, during the experiments, we found these notebooks are complicated and also include retraining models. Retraining the reinforcement learning policies would lead to a run of the tool taking 30 minutes to an hour on a laptop. Instead of executing these calls in evaluation, our bypass logic loads the corresponding pre-trained model or dataset directly from the artifacts repository. For easier reproducibility, we pre-computed and stored “artifacts” containing trained models and datasets for all of the 15 Scope-RL notebooks.


\subsubsection{Evaluation Methodology}

Our evaluation methodology differs between the two libraries to accommodate their distinct optimization objectives. For OBP, which operates on a minimization objective, we measure the relative estimation error of three primary estimators. The Direct Method (DM) estimator uses regression models to directly predict rewards, the Inverse Propensity Weighting (IPW) estimator corrects for selection bias using importance sampling, and the Doubly Robust (DR) estimator combines both approaches to achieve better bias-variance tradeoffs.

Performance improvements in OBP are indicated by negative percentage changes (error reduction), while positive values indicate degradation. The relative estimation error is calculated as the absolute difference between the estimated policy value and the ground truth, normalized by the ground truth value.

For Scope-RL, which employs a maximization objective using the relative\_policy\_value metric, positive percentage changes indicate improvement while negative values represent degradation. The library provides a comprehensive suite of estimators including basic estimators (DM, TIS, PDIS, DR, SNTIS, SNPDIS, SNDR) and marginal estimators. Values greater than 1.0 in the relative\_policy\_value metric indicate that the candidate policy outperforms the baseline policy. All performance metrics are calculated as:
\[
\text{Percentage Change} = \frac{\text{Best Iteration Value} - \text{Baseline Value}}{\text{Baseline Value}} \times 100
\]

To handle extreme optimization failures, we flag any absolute percentage change exceeding 9,999\% as "EXTREME," indicating parameter explosions rather than genuine improvements. These cases provide crucial insights into optimization stability and framework robustness.

\subsubsection{Baseline Methods}

We evaluate GrowthHacker against three baseline approaches. CrewAI represents a multi-agent framework that configures specialized AI agents for collaborative task completion. The default LLM  (using Gemini-2.5-Flash) baseline employs a standard, straightforward LLM with direct prompting for code generation without any specialized optimization framework. AutoGen~\cite{wu2023autogen}, developed by Microsoft, offers an agent-based conversational framework that allows for iterative code refinement through agent dialogue. These baselines were chosen for their widespread use and ease of adoption.


\section{Results and Discussions}
\label{sec:results}

\subsection{Overall Framework Failure Rate Analysis (RQ1)}

\paragraph{Framework-Specific Failure Patterns}

Table~\ref{tab:failure_analysis} summarizes the failure distribution across frameworks, revealing distinct vulnerability profiles. A ``failure'' refers to cases where the optimized code either fails to execute due to syntax errors, produces runtime errors, or encounters infrastructure issues that prevent successful completion. \textcolor{blue}{As mentioned earlier, the first 4 frameworks in Table~\ref{tab:failure_analysis}, Default, CrewAI, AutoGen, two\_agent(equal), uses a balanced design in which each framework--method combination receives equal replication: we tested each of the four frameworks across 162 runs (18 notebooks $\times$ 3 repetitions $\times$ 3 modification methods). }

\begin{table}[htbp]
\centering
\caption{\textcolor{blue}{Framework Failure Analysis Summary}}
\label{tab:failure_analysis}
\begin{tabular}{lcccl}
\hline
\textbf{Framework} & \textbf{Failure Rate} & \textbf{Failed Runs} & \textbf{Success Rate} & \textbf{Most Common Issue} \\
\hline
Default & 10.5\% & 17/162 & 89.5\% & Parameter placement errors \\
CrewAI & 9.9\% & 16/162 & 90.1\% & Syntax errors \& duplicates \\
AutoGen & 7.4\% & 12/162 & 92.6\% & File corruption patterns \\
two\_agent (equal) & 1.9\% & 3/162 & 98.1\% & Scope-RL runtime errors \\
two\_agent & 0\% & 0/18 & 100\% & None \\

\hline
\end{tabular}
\end{table}

\textcolor{blue}{The two\_agent (equal) framework recorded the lowest failure rate (1.9\%, 3 of 162 runs), with all three failures on Scope-RL notebooks (Basic and REC families) due to runtime parameter incompatibilities rather than syntax errors. Notably, two\_agent (equal) framework, with selected configuration, recieved 100\% Success Rate.} Contrary to initial expectations, AutoGen achieved the second-best performance with only 7.4\% failure rate, despite its sophisticated multi-agent architecture. The single-LLM default approach exhibited the highest failure rate at 10.5\%, primarily due to parameter placement errors. The concentration of failures in specific framework-method combinations provides actionable insights: AutoGen's whole\_code mode accounted for 50\% of its failures (6 out of 12), while the manual\_patch method showed the highest failure rates across default (7 failures) and CrewAI (7 failures) frameworks. This pattern suggests that framework robustness varies significantly by code application method.

\paragraph{Failure Mode Analysis}

Our comprehensive failure analysis identified four distinct failure modes that account for all failed runs across the frameworks. These failure patterns exhibit remarkable consistency, enabling predictive identification of high-risk scenarios, as shown in Table \ref{tab:failure_types}.

\begin{table}[htbp]
\centering
\caption{Failure Type Analysis and Root Causes}
\label{tab:failure_types}
\begin{tabular}{lcll}
\hline
\textbf{Failure Type} & \textbf{Frequency} & \textbf{Primary Framework} & \textbf{Root Cause} \\
\hline
Syntax \& Code Errors & 62\% of failures & AutoGen/CrewAI & Invalid parameter placement \\
File Corruption & 18\% of failures & AutoGen & Diff syntax injection \\
Infrastructure Issues & 13\% of failures & All frameworks & Google API connectivity \\
Valid Code/Runtime & 7\% of failures & Mixed & Parameter compatibility \\
\hline
\end{tabular}
\end{table}

\textbf{Syntax \& Code Errors (62\% of failures):} The most prevalent failure mode, accounting for 30 cases, involves incorrect placement of parameters in function calls and constructors. This primarily affected AutoGen and CrewAI frameworks. A representative example shows the gamma parameter incorrectly placed outside the CreateOPEInput constructor:

\begin{verbatim}
# Corrupted output example
    },
+    gamma=0.99, # ← INCORRECT: gamma parameter in invalid location
    state_scaler=MinMaxObservationScaler(
\end{verbatim}

Secondary patterns include CrewAI's systematic generation of invalid Python dictionary syntax with duplicate closing braces, suggesting challenges in maintaining code coherence across agent boundaries.

\textbf{File Corruption (18\% of failures):} The second most severe failure pattern, affecting 9 cases, involves AutoGen's whole\_code mode systematically corrupting generated files by injecting diff syntax directly into Python code:

\begin{verbatim}
# AutoGen whole_code corruption example
--- a/main.py
+++ b/main.py
@@ -148,8 +148,8 @@
-    gamma=0.95,
-    bandwidth=1.0,
+    gamma=0.99,
+    bandwidth=2.0,
\end{verbatim}

This corruption pattern is the primary driver of AutoGen's failures and accounts for the majority of file corruption cases across all frameworks (9 of 48 total failures, 18\%), causing immediate execution failure as the Python interpreter cannot parse diff syntax as code.

\textbf{Infrastructure Issues (13\% of failures):} Network infrastructure failures affected 6 cases across all frameworks during heavy computational loads. The characteristic error signature:

\begin{verbatim}
google.api_core.exceptions.DeadlineExceeded: 504 Deadline Exceeded
RetryError: Timeout of 600.0s exceeded
\end{verbatim}

These failures occurred during extended computation periods, suggesting that network latency compounds with task complexity.

\textbf{Code Validity and Execution Errors (7\% of failures):} The smallest category with 3 cases involves syntactically correct code that fails during execution due to parameter compatibility issues with underlying libraries. A representative example shows bandwidth parameter additions that are syntactically valid but incompatible with Scope-RL library requirements. Critical thresholds emerged from the analysis: bandwidth reductions below 1.0 consistently triggered numerical explosions, while learning rate increases above $6 \times 10^{-4}$ amplified instabilities. Without these findings, unsafe optimizations could be triggered.

\subsection{Comparative Performance of Frameworks (RQ2)}

\subsubsection{OBP Dataset Results}


The experiments on OBP dataset provide insights into framework performance on bandit learning tasks with clear minimization objectives. Unlike Scope-RL's maximization approach, OBP uses relative estimation error metrics (e.g., relative\_ee, relative\_ee\_ipw\_lr) where lower values represent better performance. For OBP's minimization objective, negative percentage changes indicate improvements (values decreased), while positive percentages indicate degradation (values increased).

Our methodology calculates performance changes from baseline (iteration 0) to the best iteration using ((best - baseline) / baseline) × 100, with 0.0\% assigned when baseline equals zero to avoid division errors. Results are aggregated within each notebook and policy combination, preserving policy-specific behaviors to maintain interpretability of framework-policy interactions.


\textbf{Color Legend:} \textcolor{cyan}{Cyan} indicates improvement (negative percentages), \textcolor{orange}{Orange} indicates degradation (positive percentages), and black indicates no change (0.0\%).

\paragraph{OBD Notebook Results}
The OBD notebook, utilizing real-world fashion e-commerce data from Zozotown Research, presented more stable optimization patterns. Table~\ref{tab:obd_results} shows the performance for the epsilon-greedy policy.

\begin{table}[htbp]
\centering
\caption{\textcolor{blue}{OBD Notebook Performance Results (Policy: ee). Each cell shows performance change (percentage $\pm$ standard deviation) with sample size (n=x). Negative = improvement (error reduction). }}
\label{tab:obd_results}
\begin{tabular}{lcccc}
\hline
\textbf{Estimator} & \textbf{AutoGen} & \textbf{CrewAI} & \textbf{Default} & \textbf{two\_agent (equal)} \\
\hline
DM & \textcolor{cyan}{$-10.0\%$ $\pm$ 6.2\%} \textcolor{blue}{(n=9)} & \textcolor{cyan}{$-3.9\%$ $\pm$ 5.4\%} \textcolor{blue}{(n=7)} & \textcolor{cyan}{$-2.7\%$ $\pm$ 6.0\%} \textcolor{blue}{(n=5)} & \textcolor{cyan}{$-19.8\%$ $\pm$ 10.2\%} \textcolor{blue}{(n=9)} \\
DR & $0.0\%$ $\pm$ 0.0\% \textcolor{blue}{(n=9)} & $0.0\%$ $\pm$ 0.1\% \textcolor{blue}{(n=7)} & $0.0\%$ $\pm$ 0.0\% \textcolor{blue}{(n=5)} & $0.0\%$ $\pm$ 0.0\% \textcolor{blue}{(n=9)} \\
IPW & $0.0\%$ $\pm$ 0.0\% \textcolor{blue}{(n=9)} & $0.0\%$ $\pm$ 0.0\% \textcolor{blue}{(n=7)} & $0.0\%$ $\pm$ 0.0\% \textcolor{blue}{(n=5)} & $0.0\%$ $\pm$ 0.0\% \textcolor{blue}{(n=9)} \\
\hline
\end{tabular}
\end{table}

\textcolor{blue}{The OBD results show that DM is the only estimator responsive to optimization on this dataset: two\_agent produced the largest DM reduction ($-19.8\%$), followed by AutoGen ($-10.0\%$). DR and IPW showed no change (0.0\%) across all frameworks, indicating that these estimators are insensitive to the hyperparameter changes tested.}

\paragraph{Synthetic Notebook Results}
The synthetic notebook exhibited the most extreme variations, particularly for importance-weighted estimators. Table~\ref{tab:synthetic_combined} presents the combined results for all three policies, revealing stark differences in optimization behavior across policy types.

\begin{table}[htbp]
\centering
\small
\caption{\textcolor{blue}{Synthetic Notebook Results - All Policies}}
\label{tab:synthetic_combined}
\begin{tabular}{llcccc}
\hline
\textbf{Policy} & \textbf{Estimator} & \textbf{AutoGen} & \textbf{CrewAI} & \textbf{Default} & \textbf{two\_agent (equal)} \\
\hline
\multirow{3}{*}{LR}
 & DM & $0.0\%$ $\pm$ 0.0\% \textcolor{blue}{(n=8)} & $0.0\%$ $\pm$ 0.0\% \textcolor{blue}{(n=8)} & $0.0\%$ $\pm$ 0.0\% \textcolor{blue}{(n=8)} & \textcolor{cyan}{$-1.0\%$ $\pm$ 3.0\%} \textcolor{blue}{(n=9)} \\
 & DR & \textcolor{cyan}{$-0.1\%$ $\pm$ 0.1\%} \textcolor{blue}{(n=8)} & $0.0\%$ $\pm$ 0.0\% \textcolor{blue}{(n=8)} & $0.0\%$ $\pm$ 0.1\% \textcolor{blue}{(n=8)} & \textcolor{cyan}{$-6.7\%$ $\pm$ 19.3\%} \textcolor{blue}{(n=9)} \\
 & IPW & $0.0\%$ $\pm$ 0.0\% \textcolor{blue}{(n=8)} & $0.0\%$ $\pm$ 0.0\% \textcolor{blue}{(n=8)} & $0.0\%$ $\pm$ 0.0\% \textcolor{blue}{(n=8)} & $0.0\%$ $\pm$ 0.0\% \textcolor{blue}{(n=9)} \\
\hline
\multirow{3}{*}{Random}
 & DM & $0.0\%$ $\pm$ 0.0\% \textcolor{blue}{(n=8)} & $0.0\%$ $\pm$ 0.0\% \textcolor{blue}{(n=8)} & $0.0\%$ $\pm$ 0.0\% \textcolor{blue}{(n=8)} & \textcolor{cyan}{$-3.0\%$ $\pm$ 8.8\%} \textcolor{blue}{(n=9)} \\
 & DR & \textcolor{cyan}{$-0.5\%$ $\pm$ 0.7\%} \textcolor{blue}{(n=8)} & $0.0\%$ $\pm$ 0.0\% \textcolor{blue}{(n=8)} & \textcolor{cyan}{$-0.1\%$ $\pm$ 0.3\%} \textcolor{blue}{(n=8)} & \textcolor{cyan}{$-0.5\%$ $\pm$ 0.4\%} \textcolor{blue}{(n=9)} \\
 & IPW & $0.0\%$ $\pm$ 0.0\% \textcolor{blue}{(n=8)} & $0.0\%$ $\pm$ 0.0\% \textcolor{blue}{(n=8)} & $0.0\%$ $\pm$ 0.0\% \textcolor{blue}{(n=8)} & \textcolor{cyan}{$-6.9\%$ $\pm$ 20.8\%} \textcolor{blue}{(n=9)} \\
\hline
\multirow{3}{*}{RF}
 & DM & \textcolor{cyan}{$-1.0\%$ $\pm$ 2.7\%} \textcolor{blue}{(n=8)} & \textcolor{cyan}{$-3.9\%$ $\pm$ 5.4\%} \textcolor{blue}{(n=8)} & \textcolor{cyan}{$-1.5\%$ $\pm$ 3.6\%} \textcolor{blue}{(n=8)} & \textcolor{cyan}{$-3.9\%$ $\pm$ 5.2\%} \textcolor{blue}{(n=9)} \\
 & DR & \textcolor{cyan}{$-77.3\%$ $\pm$ 14.9\%} \textcolor{blue}{(n=8)} & \textcolor{cyan}{$-53.8\%$ $\pm$ 26.2\%} \textcolor{blue}{(n=8)} & \textcolor{cyan}{$-66.7\%$ $\pm$ 30.9\%} \textcolor{blue}{(n=8)} & \textcolor{cyan}{$-46.0\%$ $\pm$ 36.2\%} \textcolor{blue}{(n=9)} \\
 & IPW & \textcolor{cyan}{$-65.4\%$ $\pm$ 28.3\%} \textcolor{blue}{(n=8)} & \textcolor{cyan}{$-48.2\%$ $\pm$ 32.7\%} \textcolor{blue}{(n=8)} & \textcolor{cyan}{$-48.9\%$ $\pm$ 33.2\%} \textcolor{blue}{(n=8)} & \textcolor{cyan}{$-27.7\%$ $\pm$ 30.2\%} \textcolor{blue}{(n=9)} \\
\hline
\end{tabular}
\end{table}

\textcolor{blue}{The combined results show policy-dependent patterns. The logistic regression (LR) and random policies were largely stable, with most frameworks producing 0.0\% change across estimators; two\_agent obtained minor reductions on LR (DM: $-1.0\%$) and random (IPW: $-6.9\%$). The random forest (RF) policy showed the largest error reductions: AutoGen reduced DR error by $77.3\%$ and IPW by $65.4\%$, while two\_agent reduced DR by $46.0\%$ and IPW by $27.7\%$. These results indicate that optimization outcomes vary by policy type, with more complex policies (RF) admitting larger reductions than simpler baselines (LR, Random).}

\subsubsection{Scope-RL Library Results}

\textcolor{blue}{We conducted 540 Scope-RL runs across 15 notebooks under the balanced design (15 notebooks $\times$ 4 frameworks $\times$ 3 methods $\times$ 3 repetitions).} The library uses a maximization objective with the relative\_policy\_value metric, where values greater than 1.0 indicate policy improvement. This evaluation revealed fundamental differences in optimization dynamics between continuous and discrete action spaces, as well as clear estimator-specific vulnerabilities.

Due to the large volume of results, we \textcolor{blue}{used} policy-averaged aggregation, pooling all per-run values across policies within each subsection to provide overall framework trends independent of specific policy choices.

\textbf{Color Legend for Scope-RL:} Since Scope-RL uses a maximization objective, \textcolor{cyan}{Cyan} indicates improvement (positive percentages), \textcolor{orange}{Orange} indicates degradation (negative percentages), black indicates no change (0.0\%), and \textcolor{red}{Red} indicates extreme values (|value| > 9,999\%).

\paragraph{Representative Discrete-Action Results}

\begin{figure*}
    \centering
    \includegraphics[width=1\textwidth]{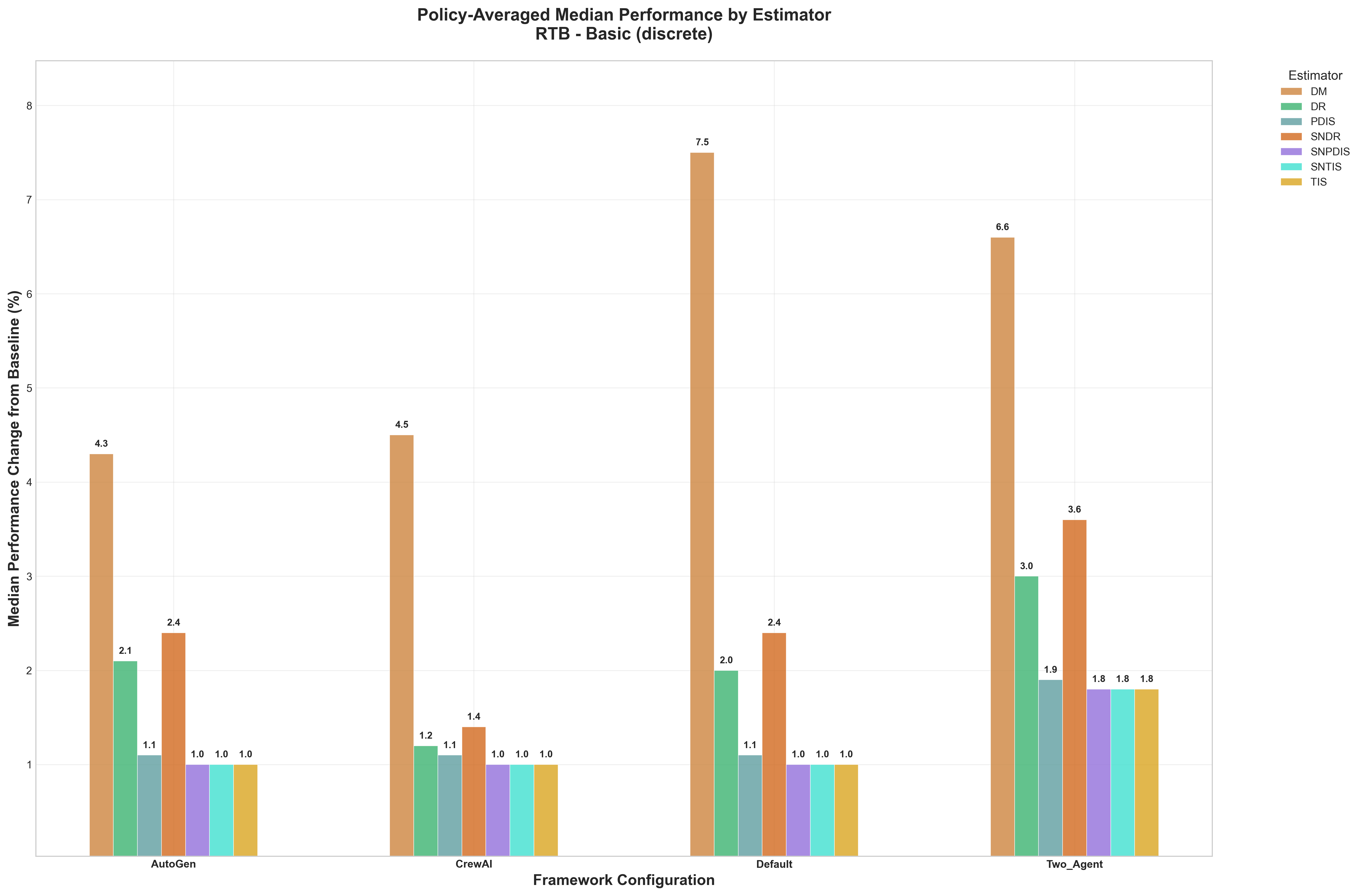}
    \caption{Visual Illustration for Performance of Estimator on RTB Basics (discrete).}
    \Description{A performance comparison figure for estimators on the RTB Basics discrete setting, contrasting framework outcomes across evaluation metrics.}
    \label{fig:rtb_basics_descrete}
\end{figure*}

We present two representative discrete-action results to illustrate framework behavior. Tables~\ref{tab:rtb_advanced_discrete} and \ref{tab:rec_basic_discrete} show performance for RTB: Advanced and REC: Basic (both discrete). Figure~\ref{fig:rtb_basics_descrete} further shows a visual illustration of the performance of estimators on RTB: Basic (discrete).

\begin{table}[htbp]
\centering
\caption{\textcolor{blue}{RTB: Advanced (discrete). Each cell shows performance change (percentage $\pm$ standard deviation) with sample size (n=x). Positive = improvement.}}
\label{tab:rtb_advanced_discrete}
\begin{tabular}{lcccr}
\hline
\textbf{Estimator} & \textbf{autogen} & \textbf{crewai} & \textbf{default} & \textbf{two\_agent (equal)} \\
& \textcolor{blue}{(n=108)} & \textcolor{blue}{(n=96)} & \textcolor{blue}{(n=84)} & \textcolor{blue}{(n=108)} \\
\hline
dm & \textcolor{cyan}{+8.2\% $\pm$ 8.1\%} & \textcolor{cyan}{+8.2\% $\pm$ 8.4\%} & \textcolor{cyan}{+10.4\% $\pm$ 8.9\%} & \textcolor{cyan}{+13.5\% $\pm$ 51.6\%} \\
dr & \textcolor{cyan}{+3.6\% $\pm$ 4.8\%} & \textcolor{cyan}{+2.7\% $\pm$ 3.7\%} & \textcolor{cyan}{+3.7\% $\pm$ 5.3\%} & \textcolor{cyan}{+8.0\% $\pm$ 16.7\%} \\
on\_policy & 0.0\% $\pm$ 0.0\% & 0.0\% $\pm$ 0.0\% & 0.0\% $\pm$ 0.0\% & 0.0\% $\pm$ 0.0\% \\
pdis & \textcolor{cyan}{+1.1\% $\pm$ 0.5\%} & \textcolor{cyan}{+1.1\% $\pm$ 0.5\%} & \textcolor{cyan}{+1.3\% $\pm$ 0.3\%} & \textcolor{cyan}{+4.5\% $\pm$ 3.0\%} \\
sndr & \textcolor{cyan}{+4.6\% $\pm$ 10.3\%} & \textcolor{cyan}{+4.2\% $\pm$ 10.4\%} & \textcolor{cyan}{+2.8\% $\pm$ 3.7\%} & \textcolor{cyan}{+7.3\% $\pm$ 17.3\%} \\
snpdis & \textcolor{cyan}{+0.9\% $\pm$ 0.5\%} & \textcolor{cyan}{+0.9\% $\pm$ 0.5\%} & \textcolor{cyan}{+1.0\% $\pm$ 0.4\%} & \textcolor{cyan}{+4.3\% $\pm$ 2.9\%} \\
sntis & \textcolor{cyan}{+0.7\% $\pm$ 0.5\%} & \textcolor{cyan}{+0.7\% $\pm$ 0.5\%} & \textcolor{cyan}{+0.8\% $\pm$ 0.5\%} & \textcolor{cyan}{+3.8\% $\pm$ 3.1\%} \\
tis & \textcolor{cyan}{+0.7\% $\pm$ 0.5\%} & \textcolor{cyan}{+0.7\% $\pm$ 0.5\%} & \textcolor{cyan}{+0.8\% $\pm$ 0.5\%} & \textcolor{cyan}{+4.4\% $\pm$ 5.7\%} \\
\hline
\multicolumn{5}{l}{\textcolor{blue}{Total=3033 (values pooled across all policies)}} \\
\hline
\end{tabular}
\end{table}

\begin{table}[htbp]
\centering
\caption{\textcolor{blue}{REC: Basic (discrete)}}
\label{tab:rec_basic_discrete}
\begin{tabular}{lcccr}
\hline
\textbf{Estimator} & \textbf{autogen} & \textbf{crewai} & \textbf{default} & \textbf{two\_agent (equal)} \\
& \textcolor{blue}{(n=16)} & \textcolor{blue}{(n=18)} & \textcolor{blue}{(n=16)} & \textcolor{blue}{(n=18)} \\
\hline
dm & \textcolor{cyan}{+10.9\% $\pm$ 13.5\%} & \textcolor{cyan}{+3.4\% $\pm$ 8.2\%} & \textcolor{cyan}{+5.5\% $\pm$ 14.0\%} & \textcolor{cyan}{+10.7\% $\pm$ 18.2\%} \\
dr & \textcolor{orange}{$-460.2\%$ $\pm$ 1225.4\%} & \textcolor{cyan}{+1922.9\% $\pm$ 8131.1\%} & \textcolor{orange}{$-136.5\%$ $\pm$ 291.6\%} & \textcolor{orange}{$-73.0\%$ $\pm$ 421.8\%} \\
on\_policy & 0.0\% $\pm$ 0.0\% & 0.0\% $\pm$ 0.0\% & 0.0\% $\pm$ 0.0\% & 0.0\% $\pm$ 0.0\% \\
pdis & \textcolor{cyan}{+1.3\% $\pm$ 0.7\%} & \textcolor{cyan}{+1.5\% $\pm$ 0.5\%} & \textcolor{cyan}{+1.3\% $\pm$ 0.7\%} & \textcolor{cyan}{+2.4\% $\pm$ 3.5\%} \\
sndr & \textcolor{cyan}{+7.9\% $\pm$ 12.7\%} & \textcolor{cyan}{+7.8\% $\pm$ 16.3\%} & \textcolor{orange}{$-1.2\%$ $\pm$ 26.3\%} & \textcolor{orange}{$-2.0\%$ $\pm$ 36.4\%} \\
snpdis & \textcolor{cyan}{+1.2\% $\pm$ 0.7\%} & \textcolor{cyan}{+1.4\% $\pm$ 0.5\%} & \textcolor{cyan}{+1.2\% $\pm$ 0.7\%} & \textcolor{cyan}{+2.1\% $\pm$ 2.9\%} \\
sntis & \textcolor{cyan}{+1.3\% $\pm$ 0.8\%} & \textcolor{cyan}{+1.5\% $\pm$ 0.6\%} & \textcolor{cyan}{+1.3\% $\pm$ 0.8\%} & \textcolor{cyan}{+2.3\% $\pm$ 3.2\%} \\
tis & \textcolor{cyan}{+1.3\% $\pm$ 0.8\%} & \textcolor{cyan}{+1.5\% $\pm$ 0.6\%} & \textcolor{cyan}{+1.3\% $\pm$ 0.8\%} & \textcolor{cyan}{+2.3\% $\pm$ 3.1\%} \\
\hline
\multicolumn{5}{l}{\textcolor{blue}{Total=535 (values pooled across all policies)}} \\
\hline
\end{tabular}
\end{table}

The results demonstrate substantial variation across environment types. In discrete action spaces (Tables~\ref{tab:rtb_advanced_discrete} and \ref{tab:rec_basic_discrete}), RTB environments show predominantly positive results across most estimators, while REC environments present more challenging optimization landscapes with substantial variance.

\subsubsection{Outcome Distribution Analysis}

\begin{figure*}
    \centering
    \includegraphics[width=1\textwidth]{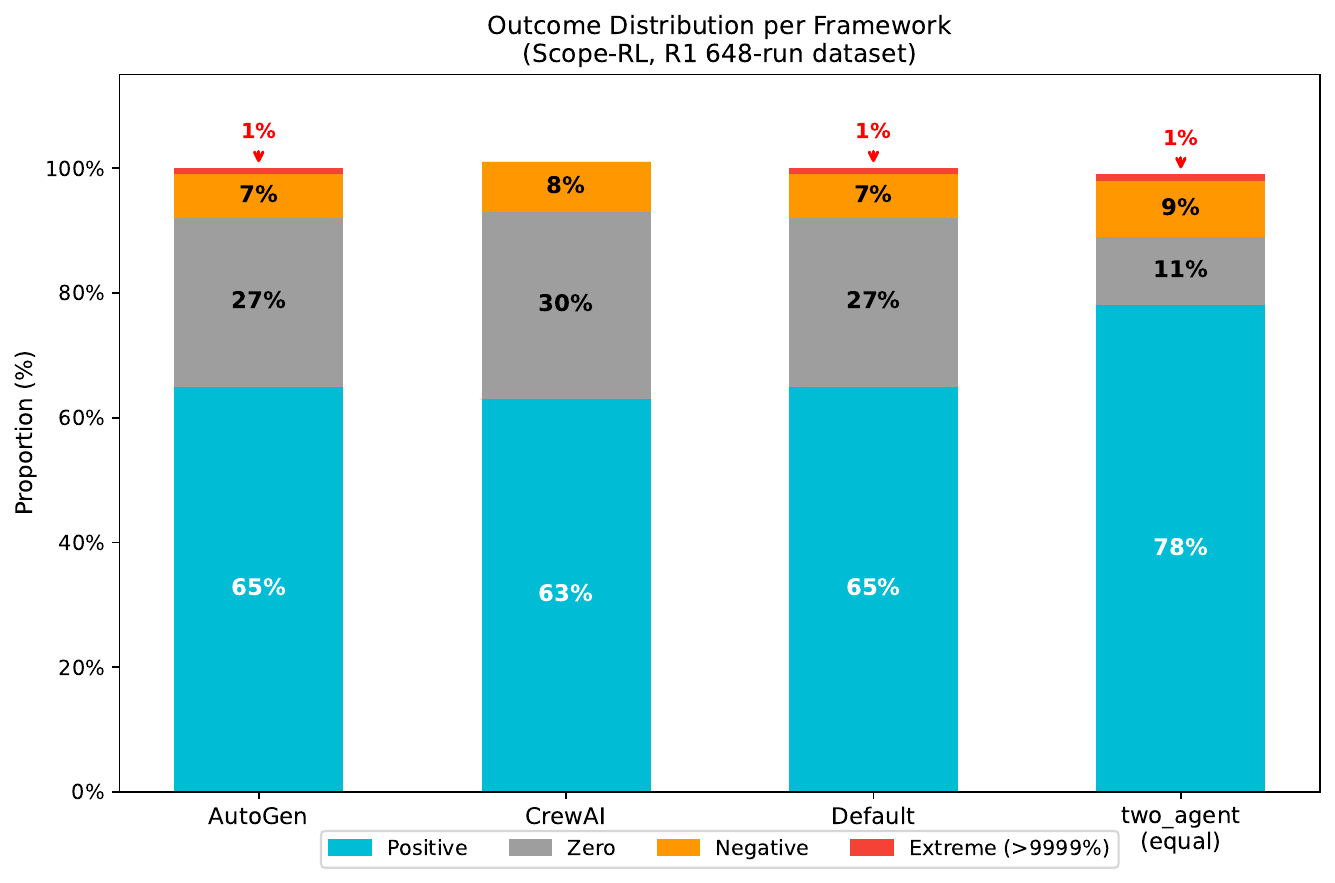}
    \caption{\textcolor{blue}{Stacked bar chart showing normalized outcome proportions per framework (Scope-RL, R1 648-run dataset). Each bar sums to 100\%.}}
    \Description{A stacked bar chart comparing the proportions of positive, neutral, negative, failed, and extreme outcomes for each framework in the Scope-RL 648-run dataset.}
\label{fig:scoperl_outcome_distribution}
\end{figure*}

Figure~\ref{fig:scoperl_outcome_distribution} presents the normalized distribution of optimization outcomes across frameworks, categorizing results into four classes: positive improvements, exactly zero change, negative degradation, and extreme values (|value| > 9,999\%).

\textcolor{blue}{The distribution analysis reveals that two\_agent achieves the highest positive outcome rate (78\%), while AutoGen and Default each achieve 65\% and CrewAI achieves 63\%. Notably, CrewAI completely avoids extreme value explosions (0\%), demonstrating superior numerical stability, while two\_agent, AutoGen, and Default maintain low extreme rates (1\% each). The zero-change proportion varies across frameworks: AutoGen and Default show approximately 27\% zero outcomes, CrewAI shows 30\%, while two\_agent shows only 11\%---reflecting two\_agent's best-iteration selection strategy, which more frequently identifies at least a marginal improvement. Negative outcome rates are consistently low across all frameworks (7--9\%).}

To quantify the magnitude of successful optimizations, Table~\ref{tab:scoperl_improvement_metrics} presents aggregate statistics for positive outcomes only.

\begin{table}[htbp]
\centering
\caption{\textcolor{blue}{Improvement Metrics for Positive Outcomes Only (Scope-RL, R1 648-run dataset)}}
\label{tab:scoperl_improvement_metrics}
\begin{tabular}{lcccc}
\hline
\textbf{Metric} & \textbf{AutoGen} & \textbf{CrewAI} & \textbf{Default} & \textbf{two\_agent} \\
\hline
Average Improvement & \textcolor{blue}{10.1\%} & \textcolor{blue}{37.9\%} & \textcolor{blue}{18.7\%} & \textcolor{blue}{26.0\%} \\
Median Improvement & \textcolor{blue}{1.7\%} & \textcolor{blue}{1.6\%} & \textcolor{blue}{1.9\%} & \textcolor{blue}{4.4\%} \\
\hline
\end{tabular}
\end{table}

\textcolor{blue}{The improvement metrics demonstrate that two\_agent achieves the highest median improvement (4.4\%) among positive outcomes, reflecting consistent gains from its best-iteration selection strategy. CrewAI achieves the highest average improvement (37.9\%), driven by occasional large-magnitude gains (e.g., in continuous action-space estimators). The substantial difference between average and median values across frameworks indicates right-skewed distributions with occasional large gains. AutoGen and Default show the most conservative profiles, with averages of 10.1\% and 18.7\% respectively and medians below 2\%.}

\textcolor{blue}{Beyond magnitude, the distribution analysis shows that even the best-performing framework (two\_agent at 78\%) leaves approximately 22\% of optimization attempts producing no change or degradation; the other frameworks show positive-outcome rates of 63--65\%. This behavior is consistent with exploratory code optimization: modifications rarely improve all estimators simultaneously, and improvements can be estimator-specific. This is acceptable in our setting because OPE (i.e., offline A/B testing) does not affect production systems or end users directly.}

\textcolor{blue}{The key differentiators are (i) two\_agent's higher positive-outcome rate (78\% vs 63--65\% for others) driven by its specialized dual-agent architecture (a dedicated Analyzer planning modifications and a Coder implementing them) and novelty-constraint tracking that avoids repeating the same suggestions across outer repetitions, and (ii) CrewAI's complete avoidance of extreme failures (0\%). The qualitative analysis in Section~5.5 illustrates the framework's reasoning using the original 7-iteration design: of the seven iterations examined, two degraded performance, but post-hoc best-iteration selection recovered a beneficial configuration.}

\subsection{Code Modification Approach Analysis (RQ3)}

To address which code modification approach yields the most reliable optimizations, we evaluated three distinct methods for applying AI-generated modifications: \textit{agent\_applies} (agent-applied partial modifications), \textit{whole\_code} (complete file replacement per iteration), and \textit{manual\_patch} (LLM-generated patches applied via Unix \texttt{patch}). Table~\ref{tab:optimization_mode_performance} presents the overall performance of each method across all experiments. \textcolor{blue}{The evaluation uses a balanced design: 18 notebooks $\times$ 3 repetitions $\times$ 4 frameworks $\times$ 3 modification methods = 648 runs, yielding 216 runs per method and 162 per framework.}

\begin{table}[htbp]
\centering
\caption{\textcolor{blue}{Optimization Mode Performance (648 runs, all four frameworks)}}
\label{tab:optimization_mode_performance}
\begin{tabular}{lccc}
\hline
\textbf{Mode} & \textbf{Success Rate} & \textbf{Runs} & \textbf{Avg Runtime} \\
\hline
agent\_applies & 93.52\% & 216 & 14m 6s \\
whole\_code & 93.06\% & 216 & 15m 6s \\
manual\_patch & 91.20\% & 216 & 13m 8s \\
\hline
\end{tabular}
\end{table}

\textcolor{blue}{All three code modification approaches yield success rates above 91\%. As Table~\ref{tab:optimization_mode_performance} shows, \textit{agent\_applies} achieved the highest success rate (93.52\%), followed by \textit{whole\_code} (93.06\%) and \textit{manual\_patch} (91.20\%). \textit{manual\_patch} is the fastest (approximately 13 minutes on average), whereas \textit{whole\_code} is the slowest (about 15 minutes). Overall, these results indicate that automated partial modifications match or slightly exceed the reliability of complete code regeneration while keeping runtimes comparable.}

The interaction between framework selection and modification method reveals compound effects on reliability, as shown in Table~\ref{tab:framework_method_combinations}.

\begin{table}[htbp]
\centering
\caption{\textcolor{blue}{Framework-Method Combination Success Rates (all frameworks evaluated across all three methods)}}
\label{tab:framework_method_combinations}
\begin{tabular}{lccc}
\hline
\textbf{Framework} & \textbf{agent\_applies} & \textbf{manual\_patch} & \textbf{whole\_code} \\
\hline
default & 92.6\% & 87.0\% & 88.9\% \\
AutoGen & 96.3\% & 92.6\% & 88.9\% \\
CrewAI & 88.9\% & 87.0\% & 94.4\% \\
two\_agent (equal) & 96.3\% & 98.1\% & 100.0\% \\
\hline
\end{tabular}
\end{table}

\textcolor{blue}{The two\_agent framework shows the highest success rate across all three modification methods (96.3\%, 98.1\%, and 100.0\% for \textit{agent\_applies}, \textit{manual\_patch}, and \textit{whole\_code} respectively), reaching 100.0\% on \textit{whole\_code}. Among the remaining frameworks, AutoGen + \textit{agent\_applies} achieves the highest success rate (96.3\%), while CrewAI + \textit{whole\_code} is the most reliable complete replacement strategy (94.4\%).} Notably, the performance ranking of modification methods varies substantially across frameworks: AutoGen performs best with \textit{agent\_applies} (96.3\%), CrewAI excels with \textit{whole\_code} (94.4\%), and the default framework shows strongest performance with \textit{agent\_applies} (92.6\%). The lowest-performing combinations were CrewAI + \textit{manual\_patch} and default + \textit{manual\_patch}, both at 87.0\%. These results indicate that framework architectural choices interact significantly with code modification strategies, and optimal pairings depend on the specific framework employed.

\subsection{\textcolor{blue}{two\_agent vs Traditional HPO and Prompt Variants (RQ4)}}

\textcolor{blue}{RQ4 addresses two questions: (1) how does the two\_agent framework compare with traditional (non-LLM) hyperparameter optimization, and (2) how do different prompt variants affect optimization outcomes? We address these through a Random Search baseline comparison and a controlled prompt-variant experiment.}

\subsubsection{\textcolor{blue}{Traditional HPO Baseline (Random Search)}}
\label{sec:baseline_random_search}

\textcolor{blue}{\paragraph{Motivation.}
To provide a conventional hyperparameter optimization (HPO) reference point, we include a safety-bounded \emph{Random Search} baseline that samples configurations uniformly within the same safety bounds used by the LLM frameworks and evaluates each by running the corresponding benchmark notebook. Unlike the LLM frameworks, Random Search does not discover tunable parameters from natural-language reasoning; instead, the set of parameters to tune is hardcoded in our bounds dictionaries, and the effective tuning scope is limited to whichever parameters the code analysis can detect in the specific code variant being optimized.}

\textcolor{blue}{\paragraph{Completion summary.}
Each baseline run corresponds to one notebook configuration (OBP: 3; Scope-RL: 15; 18 total). We label each run as: (i) \emph{passed totally} (all iterations completed), (ii) \emph{passed partially} (some iterations completed), or (iii) \emph{failed totally} (no valid output). Of 18 runs, 10 (55.6\%) produced at least one valid iteration: 5 completed all iterations and 5 completed partially, yielding 33 valid iterations in total. The remaining 8 runs failed entirely. Table~\ref{tab:random_search_baseline} shows the breakdown by library. We attribute most failures to runtime or infrastructure errors caused by blindly sampled (yet in-bounds) configurations that violate implicit library constraints.}

\begin{table}[htbp]
\centering
\caption{\textcolor{blue}{Random Search baseline completion summary.}}
\label{tab:random_search_baseline}
\begin{tabular}{lccc}
\hline
\textbf{Library} & \textbf{Passed Totally} & \textbf{Passed Partially} & \textbf{Failed Totally} \\
\hline
OBP & 3 & 0 & 0 \\
Scope-RL & 2 & 5 & 8 \\
\hline
Overall & 5 & 5 & 8 \\
\hline
\end{tabular}
\end{table}

\textcolor{blue}{\paragraph{Implications for comparison with LLM agents.}
This baseline exposes a practical limitation of conventional HPO on complex benchmarks: many runs fail to complete, wasting compute budget. All baseline failures occur in Scope-RL (8/15 failed entirely; 5/15 partial), limiting the number of evaluated configurations. In contrast, the LLM-agent frameworks in Table~\ref{tab:failure_analysis} complete 89.5\%--98.1\% of runs.}

\textcolor{blue}{Table~\ref{tab:obp_rs_vs_llm} compares Random Search with each LLM framework on the two OBP notebooks for which both approaches have complete data. Bold entries mark the largest (most negative) mean change per row. LLM frameworks yield lower error than Random Search on most estimator-notebook combinations: CrewAI produces the largest multiclass/DM reduction ($-51.6\%$) compared with Random Search's $0.0\%$, and all LLM frameworks reduce DR error ($-20.5\%$ to $-35.7\%$) whereas Random Search reaches $-83.6\%$. On the obd notebook, two\_agent produces the largest DM reduction ($-19.8\%$), while DR and IPW show $0.0\%$ change across all approaches, indicating these estimators are unresponsive to the hyperparameter changes tested.}

\begin{table}[htbp]
\centering
\caption{\textcolor{blue}{Random Search vs.\ LLM frameworks on OBP (policy: ee). Values are mean percentage change; negative $=$ improvement. \textbf{Bold} $=$ largest reduction per row.}}
\label{tab:obp_rs_vs_llm}
\begin{tabular}{llccccc}
\hline
\textbf{Notebook} & \textbf{Estimator} & \textbf{Rand.\ Search} & \textbf{AutoGen} & \textbf{CrewAI} & \textbf{Default} & \textbf{two\_agent(equal)} \\
\hline
\multirow{3}{*}{multiclass} & DM & $0.0\%$ & \textcolor{cyan}{$-31.6\%$} & \textcolor{cyan}{$\mathbf{-51.6\%}$} & \textcolor{cyan}{$-41.4\%$} & \textcolor{cyan}{$-15.2\%$} \\
 & DR & \textcolor{cyan}{$\mathbf{-83.6\%}$} & \textcolor{cyan}{$-32.3\%$} & \textcolor{cyan}{$-20.5\%$} & \textcolor{cyan}{$-35.7\%$} & \textcolor{cyan}{$-23.7\%$} \\
 & IPW & $0.0\%$ & \textcolor{cyan}{$\mathbf{-13.4\%}$} & \textcolor{cyan}{$-11.6\%$} & \textcolor{cyan}{$-10.6\%$} & \textcolor{cyan}{$-6.3\%$} \\
\hline
\multirow{3}{*}{obd} & DM & \textcolor{cyan}{$-3.3\%$} & \textcolor{cyan}{$-10.0\%$} & \textcolor{cyan}{$-3.9\%$} & \textcolor{cyan}{$-2.7\%$} & \textcolor{cyan}{$\mathbf{-19.8\%}$} \\
 & DR & \textcolor{cyan}{$\mathbf{-8.1\%}$} & $0.0\%$ & $0.0\%$ & $0.0\%$ & $0.0\%$ \\
 & IPW & $\mathbf{0.0\%}$ & $0.0\%$ & $0.0\%$ & $0.0\%$ & $0.0\%$ \\
\hline
\end{tabular}
\end{table}

\textcolor{blue}{On Scope-RL, Random Search shows higher instability than any LLM framework. In the Basic continuous family, Random Search produces a DR change of $+2{,}679\%$; in the REC Zoo discrete family, SNDR falls to $-3{,}040\%$ and SAM\_SNDR to $-4{,}520\%$, because blind sampling in high-dimensional action spaces lacks code-level safeguards. LLM agents avoid most such explosions by reasoning over estimator--parameter interactions. Of the 15 Scope-RL notebooks, 8 yield no valid Random Search optimization (only the baseline iteration completes), whereas LLM-agent frameworks complete $89.5\%$--$98.1\%$ of runs (Table~\ref{tab:failure_analysis}).}

\begin{table}[htbp]
\centering
\caption{\textcolor{blue}{Summary comparison: Random Search vs.\ LLM-based frameworks across evaluation dimensions.}}
\label{tab:rs_vs_llm_summary}
\begin{tabular}{lcc}
\hline
\textbf{Dimension} & \textbf{Random Search} & \textbf{LLM Frameworks} \\
\hline
Run completion rate & $55.6\%$\ (10/18) & $89.5\%$--$98.1\%$ \\
Params.\ modified per iteration & 2 (fixed, uniform) & Variable (code-level) \\
Best OBP improvement & DR: $-85.9\%$ & DR: $-77.3\%$ \\
Scope-RL instability events & 2 families (Basic, REC) & Reduced by code reasoning \\
Cross-estimator reasoning & None & Yes \\
\hline
\end{tabular}
\end{table}

\textcolor{blue}{Together, Tables~\ref{tab:obp_rs_vs_llm} and~\ref{tab:rs_vs_llm_summary} show that LLM-based optimization recovers the gains available to traditional HPO on OBP while providing higher reliability on Scope-RL. The added value of code-level reasoning is most visible in (a) the range of estimators improved and (b) the avoidance of parameter explosions in complex action spaces.}

\subsubsection{\textcolor{blue}{Prompt-Variant Study: Hyperparameter Tuning vs Code-Level Optimization}}

\textcolor{blue}{To isolate the contribution of code-level modifications from hyperparameter tuning, we ran a controlled prompt-variant experiment on the two\_agent framework. Variant v1 instructs the agent to ``modify the hyperparameters'' (the default behavior used in the main experiments); Variant v2 directs the agent toward ``algorithmic and structural optimizations,'' including data preprocessing, model architecture, and training procedures. Both variants enforce identical safety bounds, use the same framework and model, and ran on all 18 notebooks (36 runs total, 100\% completion rate for both).}

\textcolor{blue}{Table~\ref{tab:pv_obp} compares the two variants on OBP. Multiclass and obd use the ee policy; synthetic shows the RF policy, which shows the clearest contrast. Bold marks the better result per row. On multiclass, v2 yields a $-77.6\%$ DR reduction where v1 produces no change, while v1 retains a slight edge on IPW ($-37.3\%$ vs $-33.5\%$). On obd, v2 reduces DM error more ($-28.7\%$ vs $-17.7\%$). On synthetic (RF), v1 produces a larger DR reduction ($-62.4\%$ vs $-37.3\%$), whereas v2 finds an IPW reduction ($-6.8\%$) absent in v1. The two optimization modes thus show complementary strengths, with code-level reasoning opening improvement paths that parameter tuning alone does not reach.}

\begin{table}[htbp]
\centering
\caption{\textcolor{blue}{Prompt-variant isolation on OBP: v1 (hyperparameter-focused) vs v2 (code-level). Negative $=$ improvement. \textbf{Bold} $=$ better per row. Synthetic shows RF policy only.}}
\label{tab:pv_obp}
\begin{tabular}{llcc}
\hline
\textbf{Notebook} & \textbf{Estimator} & \textbf{v1 (HP)} & \textbf{v2 (Code)} \\
\hline
\multirow{3}{*}{multiclass} & DM & $0.0\%$ & $\mathbf{-0.5\%}$ \\
 & DR & $0.0\%$ & $\mathbf{-77.6\%}$ \\
 & IPW & $\mathbf{-37.3\%}$ & $-33.5\%$ \\
\hline
\multirow{3}{*}{obd} & DM & $-17.7\%$ & $\mathbf{-28.7\%}$ \\
 & DR & $0.0\%$ & $0.0\%$ \\
 & IPW & $0.0\%$ & $0.0\%$ \\
\hline
\multirow{3}{*}{synthetic} & DM & $\mathbf{-6.5\%}$ & $-0.1\%$ \\
 & DR & $\mathbf{-62.4\%}$ & $-37.3\%$ \\
 & IPW & $0.0\%$ & $\mathbf{-6.8\%}$ \\
\hline
\end{tabular}
\end{table}

\textcolor{blue}{On Scope-RL, Table~\ref{tab:pv_scoperl} summarizes selected notebook families that illustrate two principal patterns. The ``Pos.'' columns count estimators with strictly positive improvement, excluding EXTREME values ($>9{,}999\%$). Code-level optimization (v2) improves a wider range of estimators: on RTB: Zoo (continuous), v2 improves 15 of 17 estimators compared with 4 for v1, because code-level modifications to shared upstream computations propagate improvements across multiple estimators simultaneously. Hyperparameter tuning (v1) can occasionally reach higher single-estimator peaks through broad parameter exploration (for example, RTB: Advanced (continuous) yields 6 positive estimators under v1 vs 0 under v2), but at the cost of instability (DR $-311.8\%$ in the same configuration). Notably, the EXTREME instability events in Basic: Basic (continuous) (specifically DR, PDIS, and TIS) occur equally under both variants, indicating that the continuous action-space structure, not the optimization strategy, causes these explosions.}

\begin{table}[htbp]
\centering
\caption{\textcolor{blue}{Prompt-variant isolation on Scope-RL (policy-averaged). ``Pos.'' = number of estimators with strictly positive improvement (EXTREME excluded). DM shown as the most stable reference estimator. Both variants share 3 EXTREME events in Basic: Basic (continuous).}}
\label{tab:pv_scoperl}
\begin{tabular}{lllcccc}
\hline
\textbf{Family} & \textbf{Level} & \textbf{Action} & \textbf{v1 Pos.} & \textbf{v2 Pos.} & \textbf{v1 DM} & \textbf{v2 DM} \\
\hline
Basic & Basic & continuous & 2 & 4 & $+0.2\%$ & $+13.9\%$ \\
Basic & Advanced & continuous & 3 & 4 & $+3.9\%$ & $\mathbf{+35.5\%}$ \\
Basic & Advanced & discrete & 7 & 6 & $+0.5\%$ & $\mathbf{+20.7\%}$ \\
RTB & Zoo & continuous & 4 & $\mathbf{15}$ & $0.0\%$ & $+12.0\%$ \\
RTB & Advanced & continuous & $\mathbf{6}$ & 0 & $+18.3\%$ & $0.0\%$ \\
\hline
\end{tabular}
\end{table}

\textcolor{blue}{On Basic: Advanced, v2 yields DM improvements of $+35.5\%$ vs $+3.9\%$ (continuous) and $+20.7\%$ vs $+0.5\%$ (discrete), because v2 modifies the Q-function approximator architecture rather than only tuning its hyperparameters. The RTB: Zoo (continuous) breadth advantage (15 vs 4 estimators) arises because the code-level prompt enables the agent to restructure shared value-estimation logic, whose output propagates to all importance-weighted variants. Together, these results suggest that code-level reasoning yields optimization gains that are qualitatively distinct from, and often larger than, those from hyperparameter tuning alone.}

\subsection{\textcolor{blue}{Qualitative Analysis: Representative Modification Examples}}

\textcolor{blue}{To illustrate the modifications agents produce and the tradeoffs involved, we present a case study on the OBP \texttt{multiclass} notebook optimized by the two\_agent framework. The Analyzer Agent produced seven independent modification plans; the Coder Agent implemented each; and all were executed and scored. Table~\ref{tab:modification_examples} summarizes the relative estimation error for three illustrative iterations alongside the baseline. Listings~\ref{lst:iter1}--\ref{lst:iter7} show the code diffs produced by the agents.}

\begin{table}[htbp]
\centering
\caption{\textcolor{blue}{Relative estimation error (relative-ee) for selected iterations on \texttt{multiclass} (lower $=$ better). Percentage change is relative to baseline.}}
\label{tab:modification_examples}
\begin{tabular}{llccc}
\hline
\textbf{Iteration} & \textbf{Key changes} & \textbf{DM} & \textbf{DR} & \textbf{IPW} \\
\hline
Baseline & --- & 0.1022 & 0.0027 & 0.0150 \\
1 & C=5.0; RF n\_estimators=200 & 0.1031\; (+0.9\%) & 0.0159\; (\textcolor{orange}{+496\%}) & 0.0236\; (\textcolor{orange}{+57\%}) \\
4 & alpha\_b=0.7; alpha\_e removed & 0.2834\; (\textcolor{orange}{+177\%}) & 0.0146\; (\textcolor{orange}{+447\%}) & 0.0140\; (\textcolor{cyan}{$-7$\%}) \\
7 \textbf{(best)} & n\_folds=5; max\_iter=2000 & 0.1050\; (+2.7\%) & 0.0015\; (\textcolor{cyan}{$-44$\%}) & 0.0120\; (\textcolor{cyan}{$-20$\%}) \\
\hline
\end{tabular}
\end{table}

\textcolor{blue}{\textbf{Iteration 1 --- plausible reasoning, destabilized DR.} The Analyzer Agent identified three modifications: increase \texttt{max\_iter} from 1000 to 2000 for better convergence, add \texttt{C=5.0} to control regularization in the reward model, and set \texttt{n\_estimators=200}, \texttt{max\_depth=15} in the evaluation-policy classifier. Each change is individually reasonable. However, raising \texttt{C} to 5.0 reduces regularization in the reward model, amplifying the variance of reward-model residuals on which the DR correction term depends, increasing DR error from 0.0027 to 0.0159 (+496\%).}

\begin{listing}[htbp]
\caption{\textcolor{blue}{Iteration 1 diff (two\_agent, \texttt{multiclass})}}
\label{lst:iter1}
\begin{lstlisting}[language=diff]
-    base_classifier_e=RandomForest(random_state=12345),
+    base_classifier_e=RandomForest(n_estimators=200, max_depth=15,
+                                   random_state=12345),

-    base_model=LogisticRegression(random_state=12345, max_iter=1000),
+    base_model=LogisticRegression(C=5.0, random_state=12345, max_iter=2000),
\end{lstlisting}
\end{listing}

\textcolor{blue}{\textbf{Iteration 4 --- confident reasoning, incorrect conclusion.} The agent reduced \texttt{alpha\_b} from 0.8 to 0.7 and commented out \texttt{alpha\_e} with the note ``alpha\_e is not used in this section.'' In fact, \texttt{alpha\_e} controls the smoothing of the evaluation-policy action distribution: removing it reverts to the library default, shifting the action distribution and increasing DM error by 177\%. The code executes without error; the mistake is semantic, not syntactic.}

\begin{listing}[htbp]
\caption{\textcolor{blue}{Iteration 4 diff (two\_agent, \texttt{multiclass})}}
\label{lst:iter4}
\begin{lstlisting}[language=diff]
-    alpha_b=0.8,
+    alpha_b=0.7,

-    alpha_e=0.9,
+    #alpha_e=0.8,  # alpha_e is not used in this section
\end{lstlisting}
\end{listing}

\textcolor{blue}{\textbf{Iteration 7 --- conservative changes, best outcome.} The agent increased cross-fitting folds from 3 to 5, raised \texttt{max\_iter} to 2000, and added \texttt{n\_estimators=200}, \texttt{max\_depth=15} to the evaluation-policy classifier---without the \texttt{C} change that hurt iteration 1. The n\_folds increase appears to be the most impactful change: more folds reduce the in-sample bias of the reward-model predictions used by both DM and DR, lowering DR error to 0.0015 ($-44\%$) and IPW error to 0.0120 ($-20\%$). The post-hoc best-iteration selection mechanism identified this iteration over all seven candidates.}

\begin{listing}[htbp]
\caption{\textcolor{blue}{Iteration 7 diff (two\_agent, \texttt{multiclass})}}
\label{lst:iter7}
\begin{lstlisting}[language=diff]
-    base_classifier_e=RandomForest(random_state=12345),
+    base_classifier_e=RandomForest(n_estimators=200, max_depth=15,
+                                   random_state=42),

-    base_model=LogisticRegression(random_state=12345, max_iter=1000),
+    base_model=LogisticRegression(random_state=42, max_iter=2000),

-    n_folds=3,  # use 3-fold cross-fitting
+    n_folds=5,  # Increased n_folds for more robust estimation
\end{lstlisting}
\end{listing}

\textcolor{blue}{Taken together, these iterations illustrate common risk--reward trade-offs in LLM-based code optimization. Iteration~1 shows that interactions among seemingly reasonable parameter changes can harm estimator stability (here, increased DR variance). Iteration~4 demonstrates a semantic-error mode in which syntactically valid code silently changes program behavior. Iteration~7 shows how parallel exploration with post-hoc selection can mitigate these risks by discarding harmful variants and retaining beneficial ones.}

\section{Discussion}

Our \textcolor{blue}{648} experiments provide insights into the viability and limitations of automated OPE optimization.

\subsection{Success Rates and Failure Patterns (RQ1)}

The results show significant reliability differences across approaches, with specialized architectures achieving perfect success while general frameworks showed substantially lower rates. More complex frameworks did not guarantee better performance, for example, AutoGen's multi-agent architecture suffered from coordination failures where control commands contaminated generated code. This indicates that general-purpose frameworks carry architectural overhead that becomes problematic in specialized tasks.

The failure distribution reveals multiple concurrent challenges: syntax errors dominated in general frameworks, while parameter explosions in continuous action spaces represent a distinct safety concern that could lead to incorrect policy decisions.

\subsection{Framework Performance Comparison (RQ2)}


The two\_agent framework consistently produced stable improvements, achieving a 78\% positive-outcome rate with 26.0\% average improvement and 4.4\% median among positive outcomes. AutoGen's lower performance despite sophisticated architecture demonstrates that versatility can compromise effectiveness in specialized domains, achieving only 10.1\% average improvement and 1.7\% median. The relationship between modification strategies and framework capabilities indicates these design choices are interdependent and require careful matching; CrewAI leads in average improvement magnitude (37.9\%) and is the only framework with zero extreme-value outcomes, contrasting with two\_agent's superior consistency (highest median improvement at 4.4\%).

\subsection{Code Modification Reliability (RQ3)}

Agent-applied patches achieved highest success, followed by whole code generation and manual patches. However, substantial performance variation across framework-method combinations indicates no approach universally dominates. Manual patches exposed transformer limitations in line number calculation, while whole code generation introduced file corruption risks in certain frameworks. 

\subsection{\textcolor{blue}{Two\_Agent vs Traditional HPO and Prompt Variants (RQ4)}}

\textcolor{blue}{The Random Search baseline (Section~\ref{sec:baseline_random_search}) characterizes what parameter-only tuning can achieve under the same safety bounds. On OBP, Random Search can match strong single-estimator outcomes (e.g., $-83.6\%$ on multiclass/DR), but its gains are largely concentrated in DR. In contrast, LLM-based approaches can improve a broader set of estimators by leveraging code context (e.g., reducing DM and IPW errors).}

\textcolor{blue}{On Scope-RL, this difference is amplified: blind parameter sampling triggers instability (including parameter explosions) and yields substantially lower run-completion (55.6\%) than LLM frameworks (89.5\%--98.1\%). The prompt-variant experiment further suggests that v2's code-level edits can propagate improvements across multiple estimators (15/17 improved on RTB: Zoo (continuous)), whereas v1's hyperparameter-only changes have narrower effects (4/17).}

\subsection{\textcolor{blue}{Modification Scope and Prompt Variants}}
{\color{blue}
To contextualize the quantitative results, we briefly summarize the modification scope enforced by our system prompts.

\textbf{(i) What each agent is allowed to modify.} The system prompt (Section~3.4, ``Targeted Optimization Scope'') defines the permissible changes. Variant~v1 targets hyperparameter tuning, while Variant~v2 allows broader code-level optimizations (e.g., preprocessing and model-architecture choices). Both variants share the same framework-specific safety constraints: for OBP, adding standard scikit-learn constructor arguments to existing calls is permitted (e.g., adding \texttt{C} to \texttt{LogisticRegression}); for Scope-RL, new constructor kwargs and any structural refactoring are prohibited.

Both variants enforce explicit guardrails to preserve library API compatibility (e.g., no new keyword arguments to core OPE constructors, no new imports or external search frameworks, and no structural refactors). For Scope-RL, we additionally apply an AST-level validator before execution to reject disallowed edits.

\textbf{(ii) How these modifications lead to performance variation.} The prompt-variant experiment in Section~5.4.2 indicates that v1 and v2 are complementary rather than strictly ordered: v2 can unlock improvements by modifying shared upstream computations (e.g., improving many estimators simultaneously on RTB: Zoo (continuous)), while v1 can still provide competitive gains for estimators that are primarily hyperparameter-sensitive (e.g., IPW on OBP).}

\subsection{\JW{Implications for Software Engineering Practice}}

\JW{\textbf{Toward Automated, Continuous Offline A/B Testing via Code Modification. }For practitioners, our results indicate that continuous automated offline A/B testing with LLM-driven code modification is feasible and can be integrated into modern CI/CD workflows. This practice is analogous to mutation testing or fuzzing, where small, systematically generated code changes are explored to reveal meaningful behavioral differences. In this setting, LLM agents generate lightweight code variants, while automated OPE serves a role similar to integration tests, identifying cases where minor implementation changes lead to statistically significant improvements in data-driven A/B metrics~\cite{kohavi2020trustworthy}. As described in our motivating example in Section~1, our work demonstrates that LLM-based agents can substantially reduce the bottleneck of requirements-to-implementation flow, by automating the OPE optimization loop via code modifications. As with CI-based testing, this approach shifts validation earlier in the development cycle, reducing reliance on costly online experiments while preserving statistical rigor. For researchers, these findings motivate further investigation into agent architectures, optimization strategies, and reliability guarantees for automated requirements validation. More broadly, combining \textit{quantifiable }data-driven requirements with automated offline experimentation supports progress toward an automated, continuous requirements-to-implementation pipeline. }

\JW{%
\textbf{Industry Implications of Reliability.}
Although off-policy evaluation is typically not user-facing and is often conducted in offline experimental settings, failures such as parameter instability can still impose non-trivial costs on organizations. In the context of scope-limited reinforcement learning, parameter divergence can lead to increased computational overhead, prolonged development cycles, and, in regulated domains such as healthcare and finance, additional compliance scrutiny when unstable estimates yield unreliable validation outcomes. Observed failures in conversational multi-agent systems, such as AutoGen and CrewAI, may disrupt CI/CD pipelines, require manual intervention, and reduce developer confidence in automated evaluation workflows.}

\JW{The Two-Agent approach mitigates these risks by providing consistent OPE execution and by avoiding parameter instabilities observed in alternative designs. This level of reliability is important for practitioners who rely on automated validation in production-oriented environments, where failed experiments can result in wasted computational resources and delayed deployment timelines.
}

\JW{%
\textbf{Architectural Lessons for Agent-Based Development Tools.}
From a software engineering research perspective, our comparison of agent frameworks (Section~5.2) reveals important architectural principles for building reliable AI-assisted development tools. The two\_agent framework's superior reliability stems from its constrained architecture: file-based communication eliminates inter-agent coordination failures, role-labeled sequential execution reduces context accumulation, and targeted optimization scope ensures executable outputs. These design choices parallel established software engineering principles (modularity, separation of concerns, and well-defined interfaces), suggesting that successful AI-assisted development tools may require more constrained architectures than general-purpose agent frameworks. This contrasts with the trend toward increasingly autonomous and flexible agent systems, and suggests that domain-specific constraints may be essential for production-grade reliability in software engineering contexts.
}

\JW{%
\textbf{Jupyter Notebooks as Experimental Artifacts.}
Our evaluation uses Jupyter notebooks as code artifacts because they represent the primary development environment for OPE research and experimentation in both academic and industrial machine learning pipelines~\cite{pimentel2021understanding}. The OBP and Scope-RL libraries both distribute their examples and tutorials as notebooks, reflecting common practice in the data science community where notebooks serve as executable documentation for statistical analyses and model evaluations. While notebooks differ from production codebases in structure (cell-based vs. modular), the core optimization challenge of modifying estimator configurations and evaluation logic transfers directly to production systems where OPE code is typically encapsulated in configuration files and evaluation modules. The findings regarding reliability patterns, and framework comparison should generalize to production settings where similar modification scopes (hyperparameter tuning, estimator selection) are common. However, generalizing to larger-scale code refactoring tasks (e.g., architectural changes, multi-file modifications) remains an open question for future work.
}

\subsection{Limitations and Future Work}
Our work has several limitations that present opportunities for future research. First, our current methods do not enable agents to get iteration-aware feedback or history from previous runs/sessions. We chose this because in experiments, we tested adding iteration history, but the models got confused, which led to failures. Future work should investigate developing memory systems for agents to learn from iteration history across multiple optimization sessions.

From the experiments, we found another limitation regarding imbalance of exploration and exploitation~\cite{Badia2020NeverGU}. Specifically, in the OBP dataset, the frameworks may be over-dependent on exploitation, which could trap agents in local optima. In this case, more exploration will help in getting out of local optima. Conversely, in the Scope-RL dataset, more exploration and focus on improving accuracy are needed to avoid the failures and parameter explosions we saw in Scope-RL. Future work could focus on principled strategies for balancing exploration and exploitation (e.g., Bayesian optimization or curriculum-based search), rather than the current ``shooting in the dark'' approach in which parameter changes are proposed without directional feedback from prior iterations, contributing to the high variance in outcomes observed across notebooks.

Finally, due to the instability (such as parameter explosion phenomenon and failures) of agents observed in our experiments, human-in-the-loop systems could be considered, especially for high-stakes scenarios rather than pursuing full autonomy. As proper human validation may offer better practical tradeoffs, future work should investigate how to reach better balances between automation and human interference.

\subsection{Threats to Validity}
\paragraph{Internal Validity}
Several factors may influence the internal validity of our findings. The exclusive use of Gemini-2.5-Flash as the underlying language model, while ensuring consistency across experiments, limits our ability to generalize findings to other LLMs. Different models may exhibit varying strengths in code comprehension, parameter reasoning, and syntax generation. \JW{The framework's success may reflect overfitting to specific characteristics of OBP and Scope-RL datasets (e.g., their particular data distributions, policy structures, or estimator implementations) rather than general OPE optimization capabilities. The iteration budgets (up to 3 outer rounds for all frameworks, with one sequential modification step per round) were chosen based on preliminary experiments but may not be optimal for all scenarios. Furthermore, because the best-iteration selector does not check against the original baseline (it selects the best among the modified iterations only), if all modifications in a given round are worse than baseline, the next round begins from that degraded code rather than the original; this compounding degradation risk applies equally to all frameworks in n=2 and n=3 runs.} Additionally, the stochastic nature of LLM outputs also introduces variability, though our large sample sizes help mitigate this concern.

\paragraph{External Validity}
The generalizability of our results faces several constraints. Our evaluation focused specifically on Python-based OPE implementations using the OBP and Scope-RL libraries. \JW{The exclusive use of Python limits generalizability to other programming languages such as R or Julia, which may have different syntax constraints, library ecosystems, and error handling mechanisms. Selection of 3 OBP and 15 Scope-RL notebooks, while covering diverse OPE estimators and scenarios, introduces potential selection bias: notebooks were chosen based on OPE task focus and code completeness, potentially excluding edge cases or unusual implementations where automated optimization may be less tractable.} Results may not transfer to other programming languages, OPE frameworks, or reinforcement learning libraries with different architectural patterns. \JW{Furthermore, our focus on recommendation and advertising domains may limit applicability to other application areas such as healthcare or robotics.} The hyperparameter ranges and modification strategies tested represent a subset of possible optimization approaches, and alternative strategies might yield different performance patterns.

\paragraph{Construct Validity}
Our primary metrics (relative estimation error for OBP and relative policy value for Scope-RL) capture important aspects of optimization quality but may not reflect all relevant dimensions. The exclusive focus on performance metrics ignores other important factors such as code readability, computational efficiency, and maintainability. \JW{This may lead to optimizations that improve statistical performance at the expense of code quality or runtime efficiency.} Our definition of ``extreme values'' at 9,999\% threshold, while useful for identifying parameter explosions, is somewhat arbitrary and different thresholds might lead to different risk assessments. Additionally, success rate alone may not fully capture framework effectiveness, as some frameworks might produce higher-quality optimizations despite occasional failures.

\paragraph{Reliability}
The reproducibility of our experiments depends on several factors including LLM API stability, framework version consistency, and random seed management. While we documented all experimental artifacts, experimental runs, and their results, the evolving nature of LLM APIs means exact reproduction may become challenging over time. The proprietary nature of some components (particularly commercial LLM APIs) also limits full reproducibility by other researchers. \JW{However, our detailed documentation of prompts and evaluation protocols enables conceptual replication with alternative LLMs or updated API versions.}

\section{Related Work} \label{sec:related-work}

\subsection{LLM-based Program Synthesis and Repair}
\JW{Program synthesis automatically generates programs from high-level specifications~\cite{gulwani2017program}, while program repair techniques automatically fix software defects by generating patches that satisfy test suites~\cite{le2012systematic}. While these techniques focus on functional correctness measured by test case satisfaction, our work addresses a distinct challenge: optimizing statistical estimation quality in data-driven systems where correctness is measured by statistical properties rather than pass/fail criteria.} The recent advancement of Large Language Models has attracted significant interest in automated code optimization and generation tasks~\cite{chen2021evaluating, austin2021program}. Models like OpenAI Codex have demonstrated impressive capabilities in understanding and generating code across various programming languages~\cite{chen2021evaluating}. Several frameworks have been developed to leverage LLMs for code optimization, including RAPGen, which utilizes Codex in a zero-shot setting to identify and resolve inefficiencies in C\# code~\cite{mao2023rapgen}. Search-based approaches like SBLLM have combined evolutionary algorithms with LLMs to explore optimization spaces more systematically~\cite{jiang2023impact}. Multi-agent frameworks such as AutoGen~\cite{wu2023autogen} and CrewAI~\cite{barbarroxa2025paams} have further enhanced code generation capabilities by enabling collaborative problem-solving among multiple specialized agents. These frameworks demonstrate improved performance in identifying bugs and generating solutions compared to standalone LLM approaches. However, the application of LLM-based code optimization to data-driven decision-making tasks, particularly in the context of code modifications for statistical estimators in OPE, remains largely unexplored, presenting both opportunities and challenges for automated machine learning systems.

\subsection{Search-Based Software Engineering}
\JW{Search-based software engineering (SBSE) applies metaheuristic optimization to various software engineering problems~\cite{harman2012search}, including test generation, refactoring, and project scheduling. Recent work has explored combining search-based approaches with machine learning and large language models for code optimization~\cite{gao2024search}. Our two-agent framework shares SBSE's goal of automating optimization but differs in its approach: rather than using metaheuristic search over a discrete space, we leverage LLM agents' reasoning capabilities to navigate the combined space of code modifications and hyperparameter configurations, guided by statistical evaluation metrics specific to OPE.}

\subsection{Online and Offline A/B Testing and Off-Policy Evaluation}
\JW{A/B testing has become a cornerstone methodology for validating requirements and design decisions in modern software systems, with organizations conducting thousands of controlled experiments annually~\cite{kohavi2020trustworthy}. However, A/B testing faces significant practical limitations: deployment costs, user exposure risks, and extended experiment durations.} Off-policy evaluation \JW{addresses these constraints by enabling requirements validation using historical data without production deployment}~\cite{precup2000eligibility, thomas2016data}. The core objective of OPE is to estimate the performance of a candidate policy using data collected under a different behavioral policy~\cite{dudik2014doubly}. Various estimators have been developed to address the fundamental bias-variance tradeoff inherent in offline evaluation, including Inverse Propensity Weighting (IPW)~\cite{precup2000eligibility}, Direct Method (DM)~\cite{beygelzimer2009offset}, and Doubly Robust (DR) estimators~\cite{dudik2014doubly}. More sophisticated approaches such as Capped Importance Sampling (CIS)~\cite{bottou2013counterfactual} and Self-Normalized variants~\cite{swaminathan2015self} have been proposed to improve stability and reduce variance. Recent work has extended OPE techniques to offline reinforcement learning settings, developing estimators like Per-Decision Importance Sampling (PDIS)~\cite{precup2000eligibility} and State-Action Marginal estimators~\cite{liu2018breaking}. Despite these advances, OPE remains largely a manual process requiring domain expertise to select appropriate estimators and tune hyperparameters, limiting its scalability and reliability in practice.

\section{Conclusion} \label{sec:conclusion}

In this work, we demonstrate the viability of leveraging LLM-based agents for automatically optimizing off-policy evaluation via code refinement. The GrowthHacker system, named to automate growth hacking activities in data-driven experiments, addresses the gap in autonomously improving OPE performance through iterative code refinement, which opens up new opportunities for automated OPE optimization using code refinement agents. Our comprehensive evaluation across multiple agent frameworks (AutoGen, CrewAI, two\_agent, and default LLM configurations) reveals that automated code optimization can achieve meaningful improvements in estimation accuracy, with the two\_agent framework demonstrating particularly strong performance, achieving the highest reliability (98\%-100\% success rate) and the highest positive-outcome rate (78\%) and median improvement (4.4\%) among positive outcomes of the agents.\JW{The observed 98\%-100\% success rate of the two\_agent framework indicates that a high level of reliability is achievable in automated code optimization and underscores the importance of robust execution mechanisms for production environments, where failures can propagate into pipeline disruptions and increased operational costs.} The findings reveal both the promise and current limitations of this approach. 

Looking forward, more sophisticated Agentic AI-driven code refinement systems can be developed to optimize OPE, or more broadly, data-driven decision-making, toward autonomous optimization of machine learning pipelines. The potential for AI agents to serve as ``silent growth hackers,'' continuously and intelligently improving data-driven systems with very few or no manual intervention, is high. Future enhancements, including memory systems for learning from iteration history, LLM workflows with richer contextual understanding, advanced LLM architectures with better trade-off between exploration and exploitation, and improved execution efficiency, could be explored following this research direction.

\section{Data Availability} \label{sec:data-availability}
The data and scripts to reproduce our results are available in the replication package on Zenodo\footnote{\url{https://doi.org/10.5281/zenodo.17496869}} and GitHub.\footnote{\url{https://github.com/jie-jw-wu/ope-agent}}

{\catcode`\_=12\relax
\bibliographystyle{ACM-Reference-Format}
\bibliography{aaai2026}
}

\appendix

\end{document}